\documentclass[a4paper,man,natbib]{apa6}

\usepackage[english]{babel}
\usepackage[utf8x]{inputenc}
\usepackage{amsmath}
\usepackage{amssymb}
\usepackage{graphicx}
\usepackage[colorinlistoftodos]{todonotes}
\usepackage{url}
\usepackage{pifont}

\usepackage{amsthm}
\usepackage{mathtools, nccmath}
\usepackage{wrapfig}
\usepackage{comment}

\title{Discretization-Dependent Dissolution of Gliders in (Dis)Continuous Systems: Non-Platonic Self-Organization in Complex Systems}
\shorttitle{DisCo Gliders}
\author{Q. Tyrell Davis}
\affiliation{Boulder, CO, United States}

\abstract{
Many simulated complex systems that support persistent self-organizing patterns, {\itshape i.e.} gliders have a `state-plus-update' paradigm. This approach can be found in computational models of physics, continuous and neural cellular automata, residual connections in neural networks, and optimization methods like stochastic gradient descent. If the update is the output of a differential equation and modulated by a step size parameter, we have the familiar and general Euler method. Generally, a smaller step size is expected to yield more accurate results, at the expense of more computations to arrive at a desired end point. In other words, a simulation is usually expected to represent some real-world or mathematical ideal that can be approached as discretization, necessary for simulation on a digital computer, approaches zero. The concept of approaching an abstract ideal with finer and finer approximations is found in the philosophy of Platonism. In this work I investigate a phenomenon at odds with Platonism: the emergence of discretization as an essential ingredient for self-organization of persistent, mobile patterns found in some complex systems, known generally as gliders in cellular automata systems. I examine multiple systems supporting gliders that fit into the Euler method framework, including multiple approaches to continuous cellular automata and the Gray-Scott artificial chemistry system. Each of these systems yield one or more glider pattern-rule pairs that persist under specific, and sometimes quite coarse, discretization conditons, but become unstable at nominally more accurate, finer simulation conditions. These patterns (in combination with the systems they persist in) are clearly not approximations approaching an abstract ideal as discretization tends to zero, but exist on their own, somewhat baffling, terms that include the systematic errors of particular discretization regimes. I refer to these gliders as `non-Platonic'. Code for replicating or expanding on this work has been made available at \url{https://github.com/RiveSunder/DiscoGliders}. 
}

\begin{document}
\maketitle

\section{Introduction}
\label{intro}

Physics that can be described by differential equations can be simulated by computing, starting from some set of initial condtions,  the results of those equations over time. One of the simplest methods for doing so is the Euler method, in which the state of the system is modified by the result of a differential equation applied to said state, weighted by a step parameter, $h$.

\begin{equation}
    y(t+h) = y(t) + h * f(y(t))
\label{eqn:euler}
\end{equation}

where $y(t)$ represents the state of system at time $t$, $h$ is step size, and $f(y(t))$ is a differential equation $\frac{\partial y}{\partial t}$ describing the rate of change in $y$ with respect to time. In this work $*$ is used to denote scalar or element-wise multiplication and $\circledast$ will be used for convolution. 

Many complex systems that exhibit self-organization can be understood in terms of Equation \ref{eqn:euler}. For example, Lenia, a continuous cellular automata (CA) framework \citep{chan2019}, updates cell states $A(x,t)$ at $x$ cell locations at time $t$ according to Equation \ref{eqn:lenia}. 

\begin{equation}
A(x, t) = [A(x, t) + \Delta t * G(K_n \circledast A(x,t))]_0^1
\label{eqn:lenia}
\end{equation}

Where $G$ is the growth function, $K_n$ is a convolutional kernel defining a neighborhood, and $[\cdot]_0^1$ denotes a clamping function that restricts cell states to a range between 0 and 1.

Equation \ref{eqn:lenia} maps to Euler's method with $h = \Delta t$ and $\frac{\partial y}{\partial t} = f(y(x,t)) = G(K_n \circledast A(x,t))$. The specifics of the neighborhood kernel and the growth function give rise to complex emergent dynamics, including the self-organization of persistent, bio-reminiscent patterns that move and interact (`pseudorganisms'). 

The Euler method can be used as a first-order approximation of the physical laws of the universe of human residence. Likewise, we can think of CA and other complex computational systems as universes with their own physical laws, arranged and parameterized differently from our own. 

Historically, most CA systems that have been studied, including the well-known von Neumann universal constructor CA \citep{vonneumann1951, vonneumann1966} and Conway's Game of Life \citep{gardner1970}, are discrete systems. CA with continuous state values and updates have received some research interest over the decades, but simulating these on digital computers requires the `necessary evil', to quote \citep{rucker2003}, of discretization.  

In Equation \ref{eqn:euler}, the step size ($h$ or $\Delta t$) is an obvious source of discretization (and discretization error). 
We typically expect a smaller step size in the Euler method to produce more accurate results, at the expense of computing more updates. Simulation with too large a step size may yield a result with high error, unlikely to reflect the physical dynamics of the ideal system. Simple physics simulations tend to follow this expectation, and discretization by a step size that is too large may cause simulations to fail in catastrophic fashion.

\begin{figure}
\center
  \includegraphics[width=0.8\textwidth]{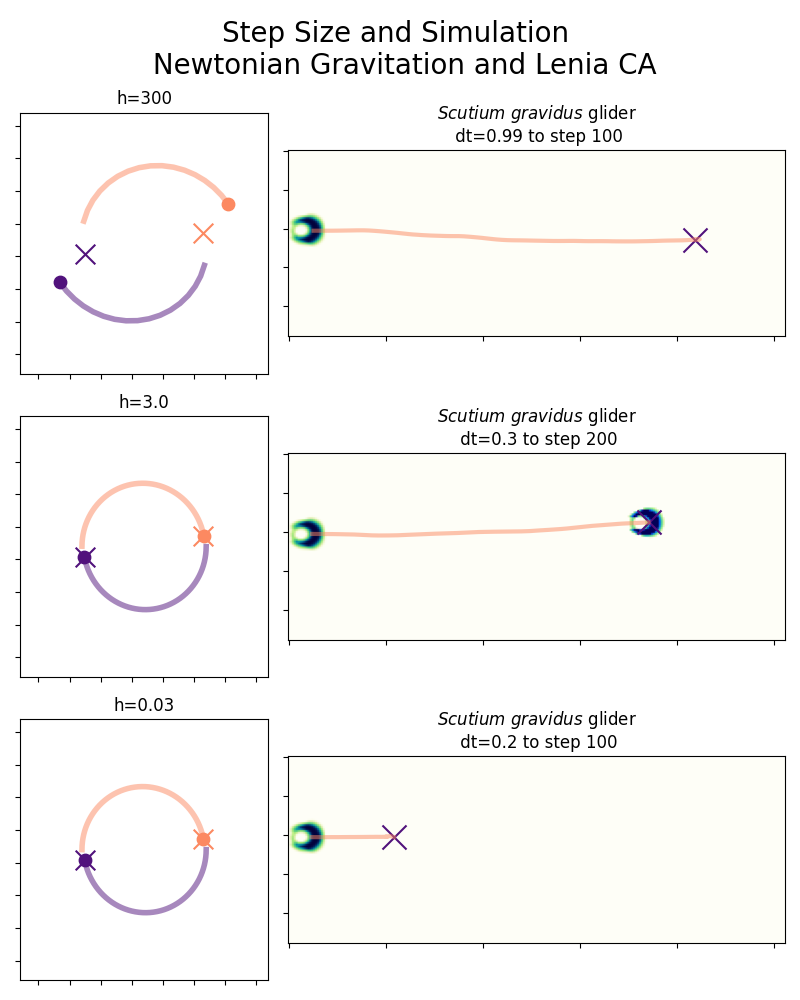}
    \caption{{\bfseries Left Column}: Gravitational attraction between two point bodies of equal mass simulated using the Euler method with different values for step size $h$. `X' markers denote the ground truth final position of each body: the result of simulating with step size $h=0.009$. Circles mark the end of each simulated path. {\bfseries Right Column}: Simulations of a glider under the {\it Scutium gravidus} update rules in the Lenia CCA framework. A very large step size $dt$ of 0.99 makes for an unstable glider. A step size of 0.3 is stable for many steps, but a slightly smaller step size of 0.2 is unstable.}
\label{fig:gravitation}
\end{figure}

Figure \ref{fig:gravitation} shows a naive simulation of gravitational dynamics using the Euler method. With too large a step size, point masses spiral out in a manner that would make a very surprising observation in our cosmos. With a step size of 3.0 or less, simulations appear to follow a consistent trajectory. Unlike the two-body gravitational example, {\it Scutium gravidus}, a CA glider found in the Lenia framework, only persists in an intermediate range of step sizes. Similar to the Newtonian gravitation example, the glider loses stability at a high step size ($\approx 0.99$). Unlike the gravitation example, the glider {\it also} loses stability in a band of much smaller step sizes from roughly 0.1 to 0.2. 

If the dynamics of a continuous CA system represent a mathematical ideal, necessarily discretized for approximate simulation, the effect of decreasing the step size should increase the accuracy of the simulation, similar to a simulation of Newtonian physics. As previously described in \citep{davis2022b} and shown in Figure \ref{fig:gravitation}, this expectation does not hold true in general. 


This work investigates self-organization in digital approximations of (nominally) continuous-valued complex systems. In particular I find several examples of glider pattern-rule pairs in a variety of complex systems where self-organization breaks down as discretization is reduced, including continuous CA, neural CA, artificial chemistry (the Gray-Scott system), and a CA based on the popular adam update method for training neural networks. 

The Related Work section discusses previous work in developing continuous CA frameworks and similar systems, with an emphasis on gliders and attention paid to discretization. 

The Methods section describes the types of complex systems and experiments I used to investigate how discretization affects self-organization of glider patterns. This includes modifications to a Gray-Scott reaction-diffusion system \citep{gray1985}, called U-Skate World \citep{munafo2014}, and a new variant of continuous CA (based on adam optimization developed by \cite{kingma2014}), that supports at least one glider. 

Results comprise the stability of glider patterns under different discretization conditions, and are presented as persistence scatter plots or a map of final grid states in a grid of discretization parameters values. The subsequent sections suggest plausible mechanisms for non-Platonic self-organization, and discuss connections between discretization, perception, and human experience.

\section{Related Work}
\label{sec:previous_work}

In one of the first implementations of continuous CA, MacLennan described a continuous spatial automaton implemented from the perspective of field computation \citep{maclennan1990}. The implementation was described for a discrete time step wherein the transition function replaces the previous cell states completely. As a consequence, MacLennan's continuous spatial automaton does not follow the state-plus-update form of Euler's method, which provides a kind of `memory' of previous cell states in the systems considered in this work: the cell states are instead replaced at each time step. No particular attention was paid to discretization effects and no gliders were described.

In discussing a 2D continuous CA framework and software package (CAPOW), Rudy Rucker described the necessity of discretization 
to simulate CA in computers as a `necessary evil' \citep{rucker2003}. While Rucker and his students did experiment with running CA simulations at a higher numerical precision (using double instead of single floating point numbers), they dismissed the modification as causing no qualitative difference in the behavior of the continuous CA, and thus not worth the added computational cost. The author did not consider gliders, but did note that in simulating the related phenomenon of waves, coarser numerical precision yielded earlier dissolution. 

The introduction of SmoothLife seems to be the first description of a glider in a continuous cellular automaton \citep{rafler2011}, though gliders in a related complex system, the Gray-Scott reaction diffusion system, were informally reported in 2009\footnote{See \url{https://www.mrob.com/pub/comp/xmorphia/uskate-world.html} and \url{https://youtu.be/ypYFUGiR51c}}. 
SmoothLife has continuous-valued neighborhoods and continuous state values, and Rafler described both a transition formulation, replacing previous cell states similar to \citep{maclennan1990}, and a state-plus-update form that fits the Euler method (Equation \ref{eqn:euler}). The glider reported in \citep{rafler2011} was discovered in the transition version of SmoothLife, with discrete time steps, but the same basic pattern is stable in the Euler update version. In considering the effect of discretization, Rafler suggested a scheme for apodization of the neighborhood kernels and to use a transition function with smooth intervals to avoid edge effects and aliasing artifacts. 

Interest in continuous cellular automata was invigorated with the description of a broad taxonomy of pseudorganisms in the Lenia framework \citep{chan2019}. Of the continuous CA systems considered in this work, CA in the Lenia framework support the largest number of described self-organizing mobile patterns\footnote{See https://chakazul.github.io/Lenia/JavaScript/Lenia.html}. Chan investigated discretization in the neighborhood kernel radius, time step, and numerical precision in bits. Observations and quantitative results in the {\itshape Orbium} glider-rule pair were consistent with the idea that, at least for {\itshape Orbium}, the CA was approximating an underlying mathematical ideal. The {\itshape Orbium} glider becomes smoother as resolution in space and time approaches zero ($\Delta t = \frac{1}{T} \to 0$, $\Delta r = \frac{1}{R} \to 0$, and precision $P \to \infty$).

In \citep{davis2022b} the authors describe a surprising departure from what had previously been considered in discretization experiments with Lenia, MacLennan's field automata, and others. The authors described observations that for some CA patterns, including a small {\it Scutium gravidus} glider described in \citep{chan2019}, stability is lost for a step size $\Delta t$ that is too small. 
The authors also demonstrated qualitative behavioral variation when modifying $\Delta t$: a `hopping' glider that exhibits a range of locomotion from straight, forward travel to following a curved `corkscrew' pattern. The results in \citep{davis2022b} were limited to two glider patterns in the {\it Scutium gravidus} Lenia CA, and to a third glider from Glaberish \citep{davis2022a}, an extended version of Lenia with the growth function split into cell-state-dependent genesis and persistence functions (analogous to birth and survival rules in Life-like CA). The Lenia framework and the Glaberish extension is described in more detail in Methods.

Motile, self-organizing patterns in complex systems (which I refer to in general as gliders), are not limited to CA. \cite{munafo2014} reported a collection of stable motile patterns in the numerical Gray-Scott reaction diffusion system \citep{gray1985}. The region of interesting behavior and glider support around a particular set of parameters is called U-Skate World, named for the archetypical and likely minimal glider in the system that resembles the letter `U'. U-Skate parameters support a variety of different glider patterns\footnote{See the U-Skate glider gallery at \url{https://www.mrob.com/pub/comp/xmorphia/catalog.html}}. Munafo was concerned about the potential effects of simulating with larger discretization values, and investigated a range of temporal step sizes from 0.5 to $\frac{1}{128} =0.0078125$ and spatial resolution variable $\Delta x$ from $\frac{1}{102}$ to $\frac{1}{572}$ (approximately $0.0098039$ to $0.0017483$). However, the spatial resolution parameter in \citep{munafo2014} modifies the the diffusion term in the Gray-Scott equations (Equations \ref{eqn:uskate_u} and \ref{eqn:uskate_v}), unlike spatial resolution in continuous CA encapsulated by the size of the neighborhood kernel. 

Munafo also compared U-Skate simulations with double and single precision floats. Double precision experiments yielded lower variance in experiments measuring glider traits like speed, but the results were otherwise qualitatively in agreement across different discretization regimes.

Munafo's work with gliders in U-Skate World built upon previous work by \citep{pearson1993}. Pearson did not find or investigate gliders, but did consider the possibility of discretization impacting simulation fidelity, reporting no qualitative difference in temporal step size from 1.0 down to 0.01 and grid sizes from 256 to 1024. 

As Lenia revitalized academic interest in continuous cellular automata, evident in many subsequent works extending, analysing, or otherwise building on \citep{chan2019}, \citep{mordvintsev2020} similarly revitalized interest in the combination of neural networks and cellular automata (Neural Cellular Automata, NCA) in a model for growing simple images from a minimal pixel seed. Like Lenia and the re-invigoration of research interest in neural systems more broadly, contemporary work on NCA is preceded by earlier work combining neural networks and cellular automata, {\itshape e.g.} \citep{wulff1992}.

In \citep{mordvintsev2020}, the authors demonstrated a type of CA with the update rule encoded in a small dense neural network shared at all locations in a 2-dimensional grid, trained via gradient descent. The NCA in \citep{mordvintsev2020} was a toy model of developmental biology (growing small images from pixel seeds), but NCA have since been used in a wide variety of scenarios including texture generation \citep{niklasson2021}, image classification/segmentation \citep{randazzo2020}, and control in a reinforcement learning context \citep{variengien2021}. 

Gliders and other self-organizing patterns in NCA have so far escaped attention in academic literature, but glider-like patterns and self-organization is evident in many NCA. For example, NCA for texture generation can yield patterns that are mobile and persist for some time (for example, see the texture NCA trained on `bubbly\_0101.jpg' in \cite{niklasson2021}). What constitutes a glider in NCA systems is not always obvious, as the boundaries between different individual patterns can be ill-defined and/or in flux, and patterns may be somewhat amorphous, transient, and variable.  

While NCA gliders have apparently not been subjected to systematic study, this seems to be due to their prevalence, rather than rarity. Discretization in NCA has also not been systematically investigated.

The seminal CA glider, Life's reflex glider, was reportedly discovered while simulating Conway's Game of Life with the stones and board for playing the game of Go \citep{berlekamp2004}. Substantial subsequent work with discrete CA gliders and related patterns has been built on that discovery, and laid the foundation for continuous CA. A review of Life and other discrete CA, especially with respect to the importance of gliders, is beyond the scope of this paper; interested readers can find background on the importance of gliders by reading Chapter 25 of Volume 4 of the ``Winning Ways'' book \citep{berlekamp2004}, and perusing the conwaylife website and forums\footnote{\url{https:://conwaylife.com}}.

\section{Methods}
\subsection{Glaberish}
\label{sec:methods}

\begin{figure}
\center
  \includegraphics[width=0.8\textwidth]{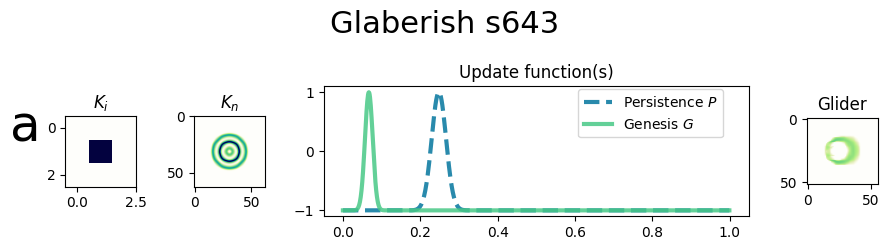}
  \includegraphics[width=0.8\textwidth]{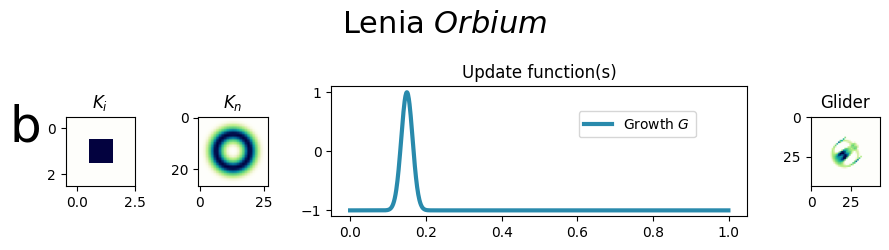}
  \includegraphics[width=0.8\textwidth]{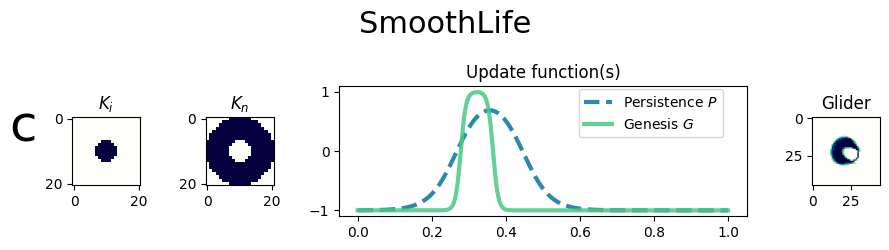}
  \includegraphics[width=0.8\textwidth]{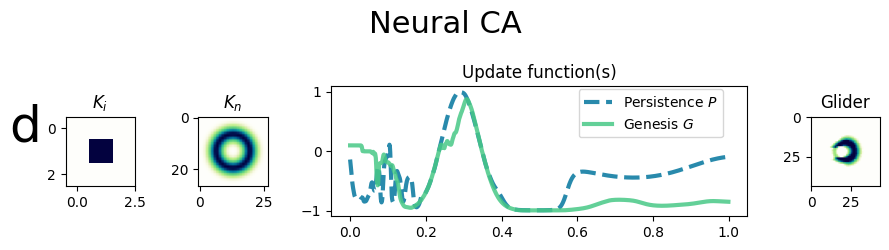}
  \includegraphics[width=0.8\textwidth]{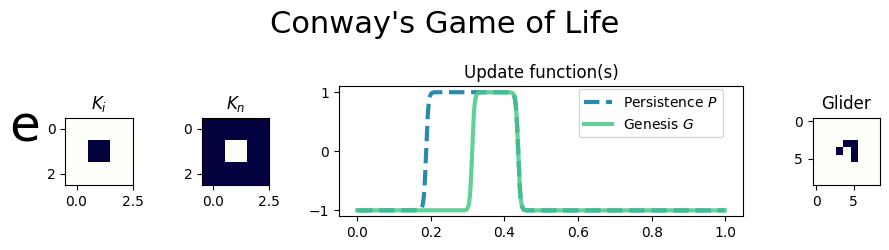}
    \caption{\bfseries{Neighborhood kernels, update functions, and typical gliders for continuous and  neural CA supporting gliders, implemented in the Glaberish framework \citep{davis2022a}.}}
\label{fig:ca_summary}
\end{figure}

This work examines the role of discretization in 6 simulated complex systems: continuous cellular automata frameworks Lenia \citep{chan2019}, SmoothLife \citep{rafler2011}, Glaberish \citep{davis2022a}, a new variant based on the adam optimization update developed by \cite{kingma2014}, and neural cellular automata. 
I also investigate a modified version of the Gray-Scott reaction diffusion system with U-Skate World parameters \citep{pearson1993, munafo2014}. 

Glaberish is a case of the Euler method where the dynamics of a continuously varying grid of cells, $A$, is described by a differential equation that takes cell states as input: $f(x) = \frac{\partial A}{\partial t}$, which I will refer to as the update equation $U$. $U$ is composed of a genesis function $G$, weighted by the difference of $1.0$ and cell states convolved with identity kernel $K_i$, and a persistence function $P$ weighted by cell states convolved with the identity kernel.

Genesis and persistence functions are analogous to the birth and survival rules defining the family of Life-like CA, Conway's Game of Life being the archetypical example. $P$ and $G$ take as input the convolution of cell states $A_t$ with neighborhood kernel $K_n$. 

Convolution kernels $K_i$ and $K_n$ can be arbitrarily defined. For example, using a scalar $1.0$ as the identity kernel and the Moore neighborhood as $K_n$, we can implement Life-like CA in Glaberish as shown in Figure \ref{fig:ca_summary}e. $P$ and $G$ can also be any function including, for example, those defined by a neural network (as in the neural cellular automata gliders considered in this work). 

Taken altogether we can write out the update function for Glaberish as the sum of the genesis and persistence functions, both acting on the result of convolving the neighborhood kernel with cell states and weighted by cell state identities. In SmoothLife, the identity kernel defines cell state identities as the convolution of cell states with a smoothing kernel ({\it i.e.} a top-hat or mesa function) with a small radius. 

\begin{equation}
\frac{\partial A}{\partial t} = U(A_t) = (1 - K_i \circledast A_t) G(K_n \circledast A_t) + (K_i \circledast A_t) P(K_n \circledast A_t)
\label{eqn:genesis_and_persistence}
\end{equation}

We can concisely describe Glaberish dynamics in terms of the update function $U$ and the state $A_t$.

\begin{equation}
A_{t+ \Delta t} = A_t + \Delta t * U(A_t) 
\label{eqn:lenia2}
\end{equation}

\subsection{Lenia}

When $U$ in Equation \ref{eqn:lenia2} only depends on neighborhoods, $K_n \circledast A_t$, we have the definition of Lenia \citep{chan2019} where $U$ is termed the growth function. The growth function for Lenia is typically a Gaussian defined by the peak center $\mu_G$ and peak width captured by $\sigma_G$, offset to yield values in the range from -1 to 1. 

\begin{equation}
G(x) = 2 {e} ^{- \frac{(x - \mu_G)^2}{2 \sigma_G^2} } - 1
\label{eqn:growth}
\end{equation}

Where $x = K_n \circledast A_t$ is the neighborhood values defined by convolution with the neighborhood kernel. The update function is shifted to a range between -1 and 1 by multiplying by amplitude 2 and subtracting 1.

Constructing Lenia neighborhood kernels also usually involves a Gaussian function. The neighborhood kernel for the archetypical Lenia CA, {\it Orbium}, has a peak value at $\mu_K=0.5$ and width parameter $\sigma_K=0.15$

\begin{equation}
K_n(r)  = a {e} ^{- \frac{(r - \mu_K)^2}{2 \sigma_K^2} }
\label{eqn:kn_lenia}
\end{equation}

and $a$ is chosen to normalize the kernel to sum to 1.0: $\frac{1}{a} = \sum_{i,j} e^{-\frac{(\sqrt{i^2+j^2} - b)^2}{c^2}}$

Update rules and the neighborhood kernel for {\it Orbium} are shown in Figure \ref{fig:ca_summary}b. The growth function has parameters $\mu_G = 0.15$, $\sigma_G=0.015$, and the neighborhood kernel has $\mu_K=0.5$ and $\sigma_K=0.15$.

\subsection{SmoothLife}

SmoothLife's transition function comprises smooth intervals, the product of sigmoid and inverted sigmoid functions. These can be described in a pair of Glaberish persistence and genesis functions. Like Conway's Life and Life-like CA, SmoothLife intervals determine which cells take on a positive value at the next time step, depending on the inner and outer neighborhood states. 

\begin{equation}
s_{\mu, \alpha}(x) = \frac{1}{1+e^{-\frac{x-\mu}{\alpha}}}
\label{eqn:smoothlife_sigmoid}
\end{equation}

\begin{equation}
G(x) = s_{\mu_{gl}, \alpha_g}(x) * (1- s_{\mu_{gu}})
\label{eqn:smoothlife_genesis}
\end{equation}

\begin{equation}
P(x) = s_{\mu_{pl}, \alpha_p}(x) * (1- s_{\mu_{pu}})
\label{eqn:smoothlife_persistence}
\end{equation}

Where $\mu_{gl}, \mu_{gu}$ and $\mu_{pl}, \mu_{pu}$ define the lower and upper edges of the genesis and persistence functions, respectively. The $\alpha$ variable modulates the smoothness of the edges. I use only one set of parameters for SmoothLife in this work: $\alpha_g=0.0280$, $\alpha_p=0.1470$, $\mu_{gl}=0.2780$, $\mu_{gu}=0.3650$, $\mu_{pl}=0.2670$, and $\mu_{pu}=0.4450$.

Unlike the other CA considered in this work, the SmoothLife identity kernel $K_i$ is not a simple 1x1 scalar kernel with value 1.0: it is defined to have non-zero values within an inner radius $r_i$, normalized to have a sum of 1.0.

\begin{equation}
K_i = \left\{ 
        \begin{array}{ c l }
            \frac{1}{M} & \quad \textrm{if } r < r_i \\
            0                 & \quad \textrm{otherwise}
            \end{array}
            \right.
\label{eqn:inner_neighborhood}
\end{equation}

The outer neighborhood kernel (Equation \ref{eqn:outer_neighborhood}) is similarly defined within the outer radius $r_o$, but is also bounded by the inner radius $r_i$ and normalized by $1/N$, $N$ being the sum of pixels between the inner and outer radii. 

\begin{equation}
K_n = \left\{ 
        \begin{array}{ c l }
            \frac{1}{N} & \quad \textrm{if } r_i \leq r < r_o\\
            0                 & \quad \textrm{otherwise}
            \end{array}
            \right.
\label{eqn:outer_neighborhood}
\end{equation}

The neighborhood kernels $K_i$ and $K_n$, update functions $G$ and $P$, and SmoothLife glider are shown in Figure \ref{fig:ca_summary}c.

\subsection{Neural CA}

The Neural CA used in this work use the same neighborhood kernel function and $\mu_K$ and $\sigma_K$ parameters as Lenia's {\itshape Orbium} rule set. 

Neural CA update rules fit within the Glaberish framework, with neural networks operating as genesis and persistence functions in Equation \ref{eqn:genesis_and_persistence}, in place of symbolic equations. The networks are trained to poorly approximate symbolic counterparts from Lenia CA. An example of neural genesis and persistence functions is shown in Figure \ref{fig:ca_summary}d. Each neural network is implemented as 3 2D convolutional layers with 1 by 1 kernel size, non-linear activation functions (Tanh or Gaussian) applied to internal layer outputs, and a dropout layer (p=0.1). The dropout layers are active only while training the networks to approximate CA functions, and turned off for stability experiments to provide deterministic output. 

A summary of the neural CA used in experiments is shown in Table \ref{tab:nca}

\begin{table}
\begin{center}
    \begin{tabular}{ | l c c c c |}
        \hline \hline
        Glider name & config & channels & activation & NN trained on \\ 
        \hline \hline
        neurorbium & neurorbium\_a  & 128 & Tanh & {\itshape Orbium} \\  
        neurosynorbium & neurosynorbium\_a & 256  & Gaussian & {\itshape Synorbium} \\  
        neurosingle & neuroscutium\_valvatus\_a & 256  & Gaussian & {\itshape Scutium valvatus} \\  
        neurowobble glider &  neuroscutium\_valvatus\_a & 256 & Gaussian & {\itshape Scutium valvatus} \\ 
        \hline
    \end{tabular}
    \caption{Summary of neural CA used in experiments}
    \label{tab:nca}
\end{center}
\end{table}

After training neural CA to approximate dynamic continuous CA, gliders were discovered using automated evolutionary methods described in \cite{davis2022c}. 





\subsection{Modified U-Skate World}

I modified U-Skate World, a particular set of parameters for the Gray-Scott reaction-diffusion system \citep{gray1985} that exhibits dynamic behavior and supports gliders \citep{pearson1993, munafo2014}. The modification involves the computational approximation of the Laplacian in the diffusion term in the Gray-Scott dynamics equations (\ref{eqn:uskate_u} and \ref{eqn:uskate_v}).

The Gray-Scott framework defines the dynamics of artificial chemical species `u' and `v'. A pair of differential equations define the change in these species over time. The first and second terms correspond to diffusion and reaction, respectively, while the third term in each equation corresponds to the feed/removal rate of u and v. The Gray-Scott equations are

\begin{equation}
\frac{\partial u}{\partial t} = \frac{1}{\Delta x^2} D_u \nabla^2 u - uv^2 + F(1-u) 
\label{eqn:uskate_u}
\end{equation}
\begin{equation}
\frac{\partial v}{\partial t} = \frac{1}{\Delta x^2}D_v \nabla^2 v + uv^2 - (F+k)v 
\label{eqn:uskate_v}
\end{equation}

Where $u$ and $v$ are the concentrations of two chemical species, $\Delta x$ is a spatial resolution scaling variable, $D_u$ and $D_v$ are the diffusion constants for $u$ and $v$, $F$ is a feed rate variable, and $k$ is the difference in the feed rates for $u$ and $v$ ($k$ is sometimes called the `decay' or `kill' rate). The reaction term $uv^2$ approximates the kinetics of a reaction that consumes one molecule of $u$, catalyzed by two molecules of $v$ to produce one additional molecule of $v$ (for a total of three).

\begin{equation}
U + 2V \rightarrow 3V
\label{eqn:reaction}
\end{equation}

The states of $u$ and $v$ are updated by the forward Euler method, as in Equation \ref{eqn:euler}. 

\begin{equation}
u_{t+\Delta t} = u_t + \Delta t * \frac{\partial u}{\partial t} 
\end{equation}
\begin{equation}
v_{t+\Delta t} = v_t + \Delta t * \frac{\partial v}{\partial t} 
\label{eqn:uskate_update}
\end{equation}

Most implementations of Gray-Scott systems use a multipoint stencil convolution kernel to approximate the Laplacian in the diffusion terms in Equations \ref{eqn:uskate_u} and \ref{eqn:uskate_v}, equivalent to the finite difference method of subtracting weighted adjacent values at each grid point. For example, one can use a five-point stencil as a convolutional kernel to approximate $\nabla^2$:

\begin{equation}
K_{D5} = \left[ 
        \begin{array}{ c c c }
            0 \hspace{1.0cm} 1 \hspace{1.0cm} 0 \\ 
            1 \hspace{0.5cm} -4 \hspace{0.9cm} 1 \\
            0 \hspace{1.0cm} 1 \hspace{1.0cm} 0 \\
        \end{array}
        \right]
\label{eqn:five_stencil}
\end{equation}

or a nine-point stencil kernel:

\begin{equation}
K_{D9} = \left[ 
        \begin{array}{ c c c }
            1 \hspace{1.0cm} 2 \hspace{1.0cm} 1 \\ 
            2 \hspace{0.35cm} -12 \hspace{0.75cm} 2 \\
            1 \hspace{1.0cm} 2 \hspace{1.0cm} 1 \\
        \end{array}
        \right]
\label{eqn:nine_stencil}
\end{equation}

Finite difference/stencil methods were used in \citep{pearson1993, munafo2014}, but my experimental objectives called for the ability to scale approximate Laplacian kernels arbitrarily, for which I used the Laplacian of Gaussian approximation 

\begin{equation}
LoG(r) = \frac{1}{\pi \sigma^2}(1 - \frac{r^2}{2\sigma^2}) e^{-\frac{r^2}{2\sigma^2}}
\label{eqn:laplacian_of_gaussian}
\end{equation}

where $\sigma$ determines the width, and $r$ is the distance from the center of the kernel at each point, normalized by division with kernel radius. I then normalize the resulting kernels to have a sum of absolute values of 1.0, as in the neighborhood kernels used elsewhere in this work.  I use a value of $\sigma=0.5$ radii when calculating the Laplacian of Gaussian kernel. The resulting 3x3 Laplacian of Gaussian kernel is

\begin{equation}
K_{log} \approx \left[ 
        \begin{array}{ c c c }
            0.0419 \hspace{1.0cm} 0.0831 \hspace{1.0cm} 0.0419 \\ 
            0.0831 \hspace{0.6cm} -0.500 \hspace{1.1cm} 0.0831 \\
            0.0419 \hspace{1.0cm} 0.0831 \hspace{1.0cm} 0.0419 \\
        \end{array}
        \right]
\label{eqn:log_stencil}
\end{equation}

which corresponds closely to the nine-point stencil method in Equation \ref{eqn:nine_stencil}, after normalizing. To support the U-Skate and Daedalus gliders from the original U-Skate world required adjusting $\Delta x$ from $\frac{1}{102} = 0.009804$ to a value of $0.00303$. The parameters used for my implementation are $\Delta x=0.00303$, $D_u =2.00 * 10^-5$, $D_v = 1.00 * 10^{-5}$, $F = 0.0620 $, $k =0.0609$, and the 3x3 kernel shown in Equation \ref{eqn:log_stencil}.

Gliders in U-Skate world were manually developed, based on patterns and common characteristics in Robert Munafo's online gallery\footnote{\url{https://www.mrob.com/pub/comp/xmorphia/catalog.html}}. The U-Skate and Daedalus patterns (Figure \ref{fig:glider_collage}r and \ref{fig:glider_collage}q) can be found in the gallery directly, and BerryCup (Figure \ref{fig:glider_collage}p) is a new pattern developed by the author.  

\subsection{Adam Automaton}

Adam optimization is a method for updating parameters, often based on gradients with respect to an objective function in large neural network models \citep{kingma2014}. 
The updates take the general form of the Euler method (Equation \ref{eqn:euler}), with the learning rate as a step size and the quotient of the first moment $m$ and the square root of the second moment $\sqrt{v}$ of the gradient (and a small variable $\epsilon$ to avoid division by 0) amounting to the differential equation describing the change in parameters. 

In equation form, the adam update is

\begin{equation}
A_{t+\Delta t} = A_t + \Delta t \frac{m}{\sqrt{v}+\epsilon}
\label{eqn:adam_update}
\end{equation}

For an adam cellular automaton, the grid states $A_t$ stand in for {\itshape e.g.} neural network parameters $\theta$ and the step size $\Delta t$ stands in for the learning rate $\alpha$.

I used the same neighborhood kernel as Lenia's {\itshape Orbium} CA, and the same Gaussian function (Equation \ref{eqn:growth} with similar parameters as a growth function).

The first moment is an exponential average of the result of the growth function applied to neighborhoods. Smoothness is determined by the exponential averaging variable $\beta_1$.  

\begin{equation}
m_{t+\Delta t} = \beta_1 m_t + (1-\beta_1) G(K_n \circledast A_t)
\label{eqn:first_moment}
\end{equation}

The second moment is likewise calculated as an exponential average of the square of the results of the growth function, with averaging variable $\beta_2$. 

\begin{equation}
v_{t+\Delta t} = \beta_2 v_t + (1-\beta_2) G(K_n \circledast A_t)^2
\label{eqn:second_moment}
\end{equation}
 
When the growth function (Equation \ref{eqn:growth}) has a value of -1.0 in the absence of neighoring cells with non-zero state, the second moment tends to a default value of $(-1)^2 = 1.0$. 

I selected and cropped sections from dynamic adam automata simulations with randomly initialized patches, eventually discovering a simple glider reminiscent of the {\itshape Orbium} glider. Based on the origins of the glider and the adam CA, I call the glider and the CA system that supports it `Adorbium'. The glider is visualized in Figure \ref{fig:glider_collage}s, and Adorbium CA parameters are $\beta_1=0.8$, $\beta_2=0.99$, $\epsilon=10^{-8}$, $\mu_G=0.167$, $\sigma_G=0.013$, $\mu_K=0.5$, $\sigma_K=0.15$, and kernel radius $k_r=13$ pixels.

\subsection{Adaptive Random Walk and Linear Grid Search}

I used a combination of adaptive random walks (for SmoothLife, Lenia, Glaberish, and neural CA) and linear grid search (for U-Skate World and Adorbium CA) to vary discretization parameters in glider persistence experiments. Random search is often preferred over grid search as a more efficient (in terms of using computational resources or wall time) means of hyperparameter tuning \citep{bergstra2012}. I used a similar approach to search over discretization parameters, starting at the native settings (that were used in the discovery of a given glider) and randomly perturbing discretization parameters within a range. 

The degree of random perturbation of discretization values depends on the difference between the persistence steps achieved for a given set of discretization parameter values and the number of persistence steps achieved in the previous iteration. In other words, my experiments were intended to make larger changes to the discretization parameter values when the current and last experimental iteration had the same or similar results (number of update steps the glider persists), and smaller changes when results were very different. The idea behind this approach was to sample more densely near boundaries in discretization parameter values supporting glider persistence. 

Succinctly, my experiments add random Normal noise to the discretization parameter values for step size $\Delta t$ and kernel radius $k_r$. Step size is then constrained to fall between a minimal value (usually 0.01) and 1.0, and kernel radius is rounded to a whole number and constrained between a minimum and maximum size.

\begin{equation}
\theta_{n+1} = \theta_n + \mathcal{N}[\theta_n, \sigma_n]
\label{eqn:biased_walk}
\end{equation}

Where $\theta_n$ is the discretization parameter value from experiment sample $n$. The standard deviation $\sigma_n$ depends on the number of persistence steps in the current sample run and the last.

 %

$\sigma_n$ is modified after each iteration by multiplication with a scaling factor $a_{\sigma}$

\begin{equation}
\sigma_{n+1} = [\sigma_n * a_{\sigma} ]_{\sigma_{min}}^{\sigma_{max}}
\label{eqn:mod1}
\end{equation}

Where $[\cdot]_{\sigma_{min}}^{\sigma_{max}}$ clamps the result between minimum and maximum values. The scaling factor is determined by 

\begin{equation}
a_{\sigma} = \left[ \left(\frac{(p_n + p_{n-1})}{(|p_{n} - p_{n-1}|+1} - 1 \right)\right]_{a_{min}}^{a_{max}}
\label{eqn:mod2}
\end{equation}

Where $p_n$ is the number of persistence steps for iteration $n$. There is no particular theoretical reason behind Equation \ref{eqn:mod2}, it was chosen empirically. Values for the scaling factor were clamped between a minimum and maximum of 0.5 and 1.5 as denoted by $[\cdot]_{a_{min}}^{a_{max}}$. 

For experiments which also modify the numerical precision, the data type (16, 32, or 64-bit floats) is sampled randomly with equal probability for each iteration. 

The CA systems used in random walk experiments lend themselves to automatic detection of glider persistence (or lack thereof). In my experiments I define persistence as the homeostatic ability for a glider pattern in a CA system to maintain a roughly consistent sum of cell states. Gliders can lose persistence in two ways: by vanishing, where the sum of cell states falls below a relative threshold; or by overgrowth, where the sum of cell states exceeds the original sum by that same relative threshold. The threshold used in my experiments is typically a relative deviation of 0.30, {\itshape e.g.} iterations are halted at the CA step where the sum of cell states, relative to the starting sum, is above $\approx 1.3$ or below $\approx 0.7$.

U-Skate and Adorbium systems have non-zero background values, making automatic persistence detection less tenable. Although I did try using correlation of the current grid with the original pattern, using a few different normalization schemes, this seemed unreliable in my hands. Instead I ran each iteration for the full duration of update steps and manually assessed the final grid state to determine whether each grid contains a glider. This limits the number of iterations both by the number that can reasonably be assessed manually (I used about 500 for each glider in U-Skate world), and because each iteration runs for the full maximum number of update steps. I used a linear grid search to get even coverage across a range of discretization values.

\section{Results}
\label{sec:results}

\begin{figure}
\center
  \includegraphics[width=0.9\textwidth]{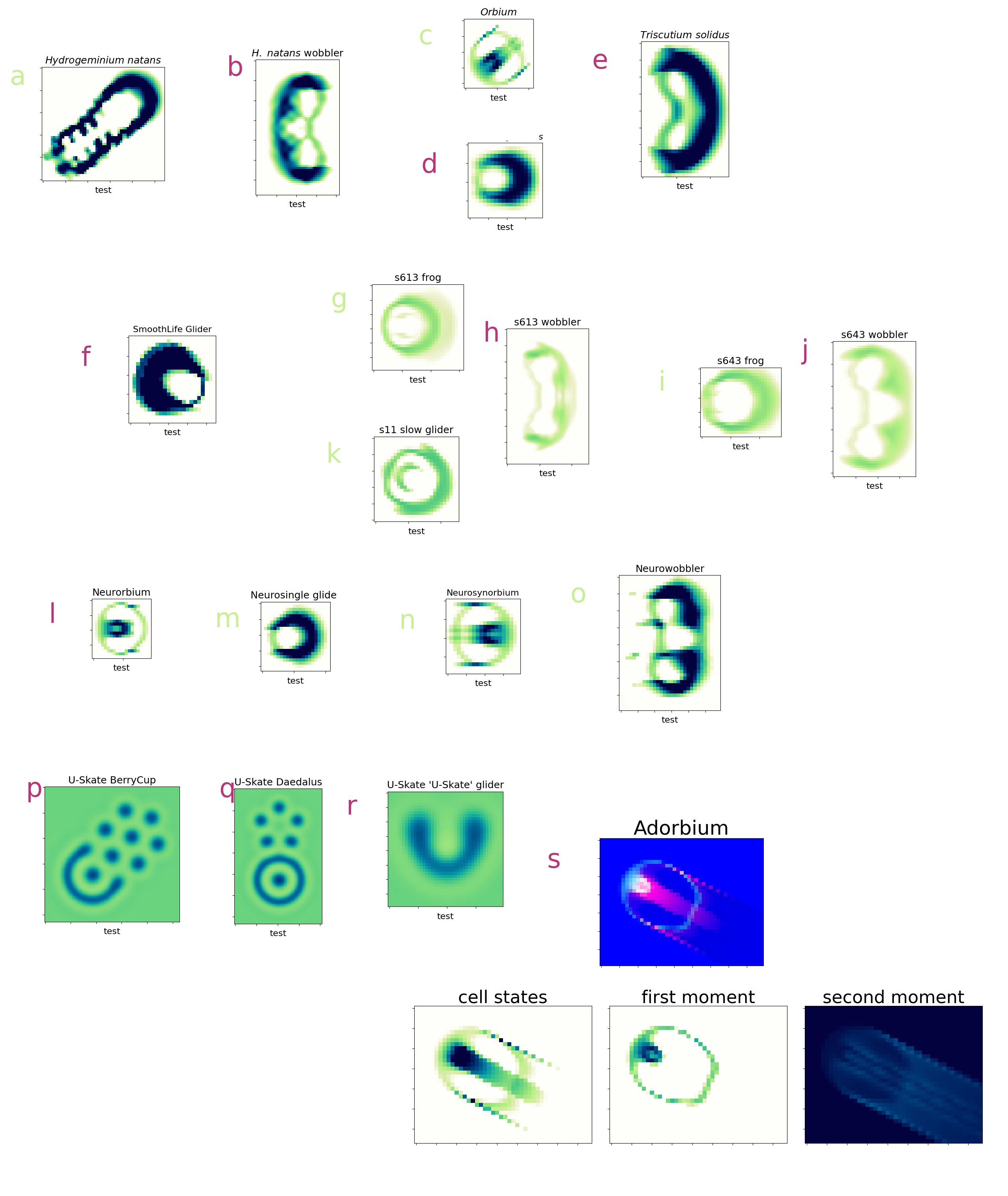}
    \caption{\small Gliders in this study. Dark red and light green letters denote non-Platonic and Platonic pattern-rule pairs, respectively. a) {\itshape Hydrogeminium natans}, Lenia, Platonic. b) {\itshape Hydrogeminium natans} wobbler, Lenia, non-Platonic. c) {\itshape Orbium}, Lenia, Platonic. d) {\itshape Scutium gravidus} glider, Lenia, non-Platonic. e) {\itshape Triscutium solidus} wide glider, Lenia, Non-Platonic. f) SmoothLife glider, SmoothLife, non-Platonic. g) s613 frog, Glaberish, Platonic. h) s613 wobbler, Glaberish, non-Platonic. i) s643 frog, Glaberish, Platonic. j) s643 wobbler, Glaberish, non-Platonic. k) s11 slow glider, Glaberish, Platonic. l) Neurorbium, NCA, non-Platonic. m) Neurosingle glider, NCA, Platonic. n) Neurosynorbium, NCA, Platonic. o) Neurowobbler, NCA, Platonic. p) BerryCup, U-Skate (Gray-Scott reaction diffusion system), putatively non-Platonic. q) Daedalus, U-Skate, putatively non-Platonic, r) `U-Skate' glider, U-Skate, putatively non-Platonic. s) `Adorbium', Adam optimizer automaton, non-Platonic.}
\label{fig:glider_collage}
\end{figure}

The investigations in this work aimed to address 3 hypotheses. The first hypothesis was that some glider pattern-rule pairs in continuous value cellular automata lose stability when discretization parameters are too fine, {\it i.e.} that systems simulated with higher nominal accuracy can lose glider stability; the second predicted that the phenomenon described in the first hypothesis is not specific to cellular automata, but can be found in multiple types of complex systems that support gliders\footnote{Furthermore, non-Platonic pattern persistence is likely a more general feature of complex systems. In the discussion section we will consider the effect of step size on attractors in the Lorenz system and the differential equation of the logistic function.}; the third hypothesis predicted that support for glider stability in a given system would be clustered around the discretization settings used in the pattern's discovery or creation. 

I categorize pattern-rule pairs as Platonic and non-Platonic. The former classification denotes systems that plausibly correspond to a mathematical ideal, while for non-Platonic systems, discretization is essential for self organization of a pattern-rule pair. A system is non-Platonic with respect to specific discretization parameters iff, for a given set of discretization parameters for which self-organization does not persist ({\itshape e.g.} the glider dissolves or explodes), there is another set of coarser discretization parameters for which the pattern is stable. For example, if a glider vanishes after a few steps in a given CA with discretization parameters $\Delta t=0.20$, and kernel radius $k_r = 17$, but can persist for thousands of steps at $\Delta t=0.225$ and $k_r = 13$, the pattern-rule pair meets the criteria of a non-Platonic system w.r.t. step and kernel size, {\itshape i.e.} temporal and spatial resolution.

To address these expectations, I considered gliders in a variety of different continuous CA frameworks while varying step size $\Delta t$, the kernel radius $k_r$, and numerical precision (float16, float32, or float64 datatypes in PyTorch). My results contained examples of glider dissolution with increasing discretization fineness (nominally greater simulation accuracy) in every type of complex system considered. 

Numerical precision can affect glider persistence as well, and my experiments included runs using \texttt{torch.float16}, \texttt{torch.float32}, and \texttt{torch.float64} data types. Dependence on floating point precision data types is not easy to discern in the persistence scatter plots. As an alternative, Table \ref{tab:dtype} contains examples of persistence dependence on data type and readers are enouraged to explore this effect using demonstration notebooks in the project's code repository: \url{https://github.com/RiveSunder/DiscoGliders}.

\subsection{SmoothLife}

The simple SmoothLife glider (Figure \ref{fig:glider_collage}f) examined in this work is based on the image and CA parameters published in \citep{rafler2011}. The glider was originally discovered in the spatially continuous, temporally discrete version of SmoothLife by Rafler, but the glider pattern was later found to stabilize in SmoothLife implemented with Euler method updates, and re-discovered by evolutionary methods \citep{davis2022c}. Under conditions similar to those described by Rafler, with a large step size near 1.0 and a kernel radius of 11 pixels, the glider is stable. 

\begin{figure}
\center
  \includegraphics[width=1.0\textwidth]{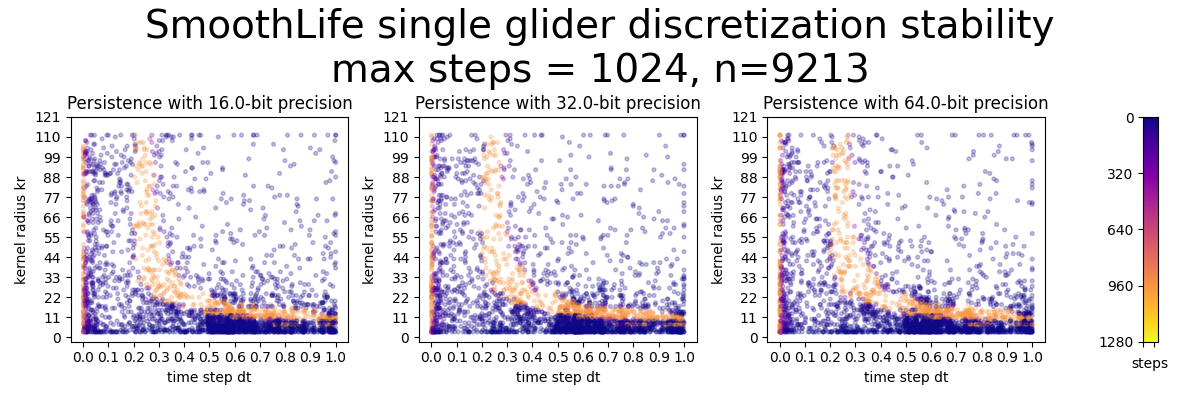}
    \caption{Stability persistence scatter plot for the SmoothLife single glider.}
\label{fig:smoothlife_map}
\end{figure}

In my experiments SmoothLife glider persistence depends on the interaction of spatial resolution (increasing with larger kernel size) and temporal resolution (the update step size $\Delta t$); it loses persistence at finer spatiotemporal resolution outside of a support region, meeting the criteria for non-Platonism w.r.t. spatiotemporal resolution. The shape of this dependence is qualitatively similar across 16, 32, and 64 bit PyTorch float datatypes, and the persistence scatter plot for the SmoothLife glider is shown in Figure \ref{fig:smoothlife_map}. The SmoothLife glider also demonstrates non-Platonic persistence with respect to numerical precision (see Table \ref{tab:dtype}).

\begin{table}
\begin{center}
    \begin{tabular}{ | l | c | c |}
    \hline \hline
Glider name & spatiotemporal discretization & torch datatype stability at 1024 steps \\ 
\hline \hline
SmoothLife & $\Delta t = 0.19$, $K_r = 36$ & float16[\ding{51}], float32[\ding{55}], float64[\ding{55}] \\  
{\itshape Scutium gravidus} & $\Delta t = 0.24$, $K_r = 13$ & float16[\ding{51}], float32[\ding{55}], float64[\ding{55}] \\  
{\itshape Triscutium solidus} & $\Delta t = 0.09$, $K_r = 11$ & float16[\ding{55}], float32[\ding{51}], float64[\ding{51}] \\  
{\itshape H. natans} wobbler & $\Delta t = 0.41$, $K_r = 19$ & float16[\ding{51}], float32[\ding{51}], float64[\ding{55}] \\  
s613 wobbler & $\Delta t = 0.066$, $K_r = 13$ & float16[\ding{51}], float32[\ding{55}], float64[\ding{55}] \\  
s643 wobbler & $\Delta t = 0.08$, $K_r = 21$ & float16[\ding{51}], float32[\ding{55}], float64[\ding{55}] \\  
Neurorbium & $\Delta t = 0.04$, $K_r = 25$ & float16[\ding{55}], float32[\ding{51}], float64[\ding{51}] \\  
Neurorbium & $\Delta t = 0.14$, $K_r = 24$ & float16[\ding{55}], float32[\ding{51}], float64[\ding{55}] \\  
    \hline
    \end{tabular}
    \caption{Examples of glider stability dependence on floating point numerical precision.}
    \label{tab:dtype}
    \end{center}
\end{table}

\subsection{Lenia}

\begin{figure}
\center
  \includegraphics[width=1.0\textwidth]{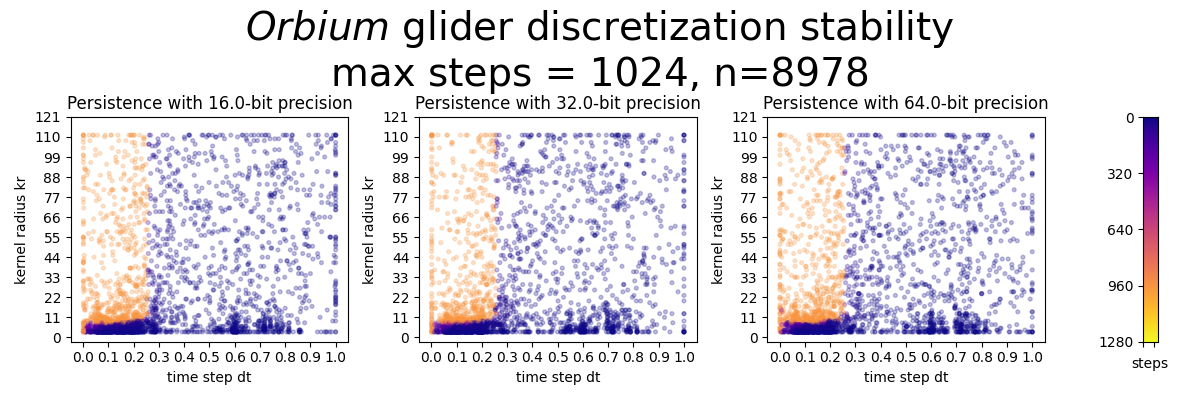}
    \caption{Stability persistence for the {\itshape Orbium} glider pattern in the Lenia CA of the same name, with $\mu_G=0.15$, $\sigma_G=0.015$, and $\mu_K=0.5$, $\sigma_K=0.15$.}
\label{fig:orbium_map}
\end{figure}

\begin{figure}
\center
  \includegraphics[width=1.0\textwidth]{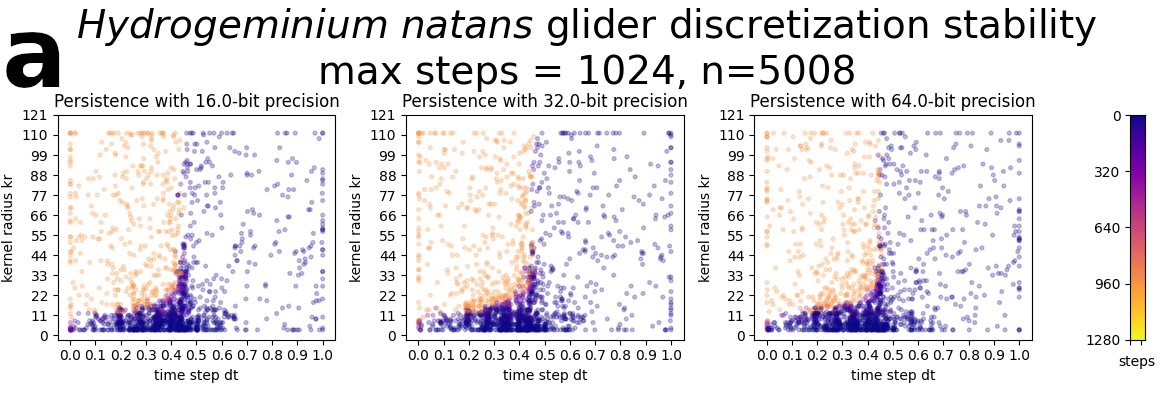}
  \includegraphics[width=1.0\textwidth]{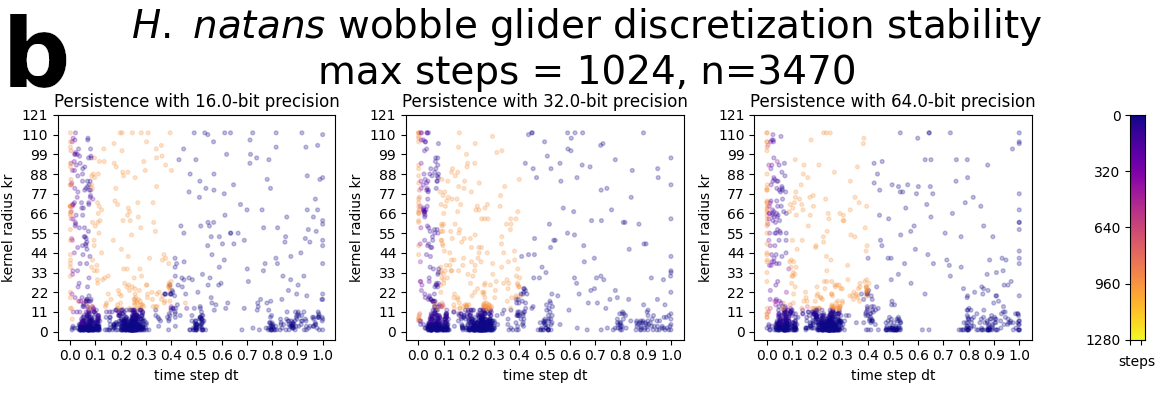}
    \caption{Stability persistence for two glider patterns in {\itshape Hydrogeminium natans}, a Lenia CA}
\label{fig:hydrogeminium_map}
\end{figure}

I examined discretization for 5 glider patterns in 4 different rule sets in the Lenia framework: 2 glider patterns in {\itshape Hydrogeminium natans} (Figure \ref{fig:glider_collage}a and b), the archetypical {\itshape Orbium} (Figure \ref{fig:glider_collage}c), {\itshape Scutium gravidus} (Figure \ref{fig:glider_collage}d), and {\itshape Triscutium solidus} (Figure \ref{fig:glider_collage}e). {\itshape Orbium} and the eponymous {\itshape Hydrogeminium natans} gliders fit the criteria for ideal, Platonic pattern-rule pairs: coarser discretization does not result in these patterns regaining persistence stability. The persistence scatter plot for {\itshape Orbium} is shown in Figure \ref{fig:orbium_map}, and for {\itshape Hydrogeminium natans} in Figure \ref{fig:hydrogeminium_map}a. 

\begin{figure}
\center
  \includegraphics[width=1.0\textwidth]{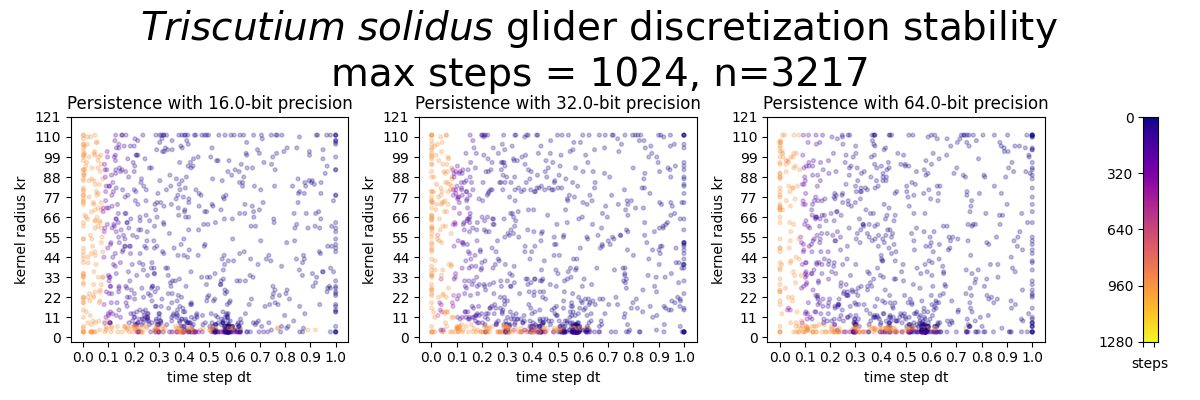}
    \caption{Stability persistence for a glider pattern in {\itshape Triscutium solidus}, a Lenia CA}
\label{fig:triscutium_map}
\end{figure}

\begin{figure}
\center
  \includegraphics[width=1.0\textwidth]{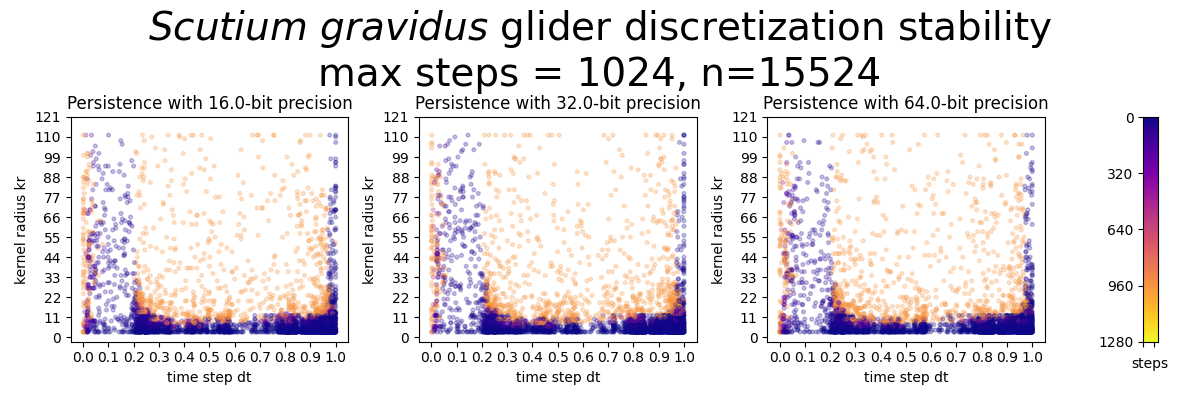}
    \caption{Stability persistence for a glider pattern in {\itshape Scutium gravidus}, a Lenia CA}
\label{fig:scutium_map}
\end{figure}

The {\itshape Triscutium solidus}, {\itshape Scutium gravidus} simple glider, and the wide wobble glider in {\itshape Hydrogeminium natans} do fit the criteria for non-Platonic pattern-rule pairs. For those gliders, under their respective CA rules, gliders may be unstable at a given combination of discretization parameters and regain stability by increasing discretization coarseness. The stability scatter plot for {\itshape Triscutium solidus} is shown in Figure \ref{fig:triscutium_map}, {\itshape Scutium gravidus} in Figure \ref{fig:scutium_map}, and the {\itshape H. natans} wobble glider in Figure \ref{fig:hydrogeminium_map}b.  

The small {\itshape Scutium gravidus} glider demonstrates non-Platonic persistence with respect to numerical precision, as does the {\itshape H. natans} wobbler glider. I did not find a specific example of non-Platonic persistence due to numerical precision for {\itshape Triscutium solidus} (see Table \ref{tab:dtype}).

\subsection{Glaberish}

Gliders and rules for Glaberish CA were evolved in \citep{davis2022c}, and include a slow glider (Figure \ref{fig:glider_collage}k), found in the rule set `s11' (named for the random seed used in the evolutionary algorithm), and two very similar `frog' gliders named for their pulsing movement patterns (Figures \ref{fig:glider_collage}g and \ref{fig:glider_collage}i) and two wobble gliders (Figures \ref{fig:glider_collage}h and \ref{fig:glider_collage}j) which are similar to the wobble glider found in Lenia's {\itshape Hydrogeminium natans} rule set. The frog and wobble gliders reside in rule sets s613 and s643.  

\begin{figure}
\center
  \includegraphics[width=1.0\textwidth]{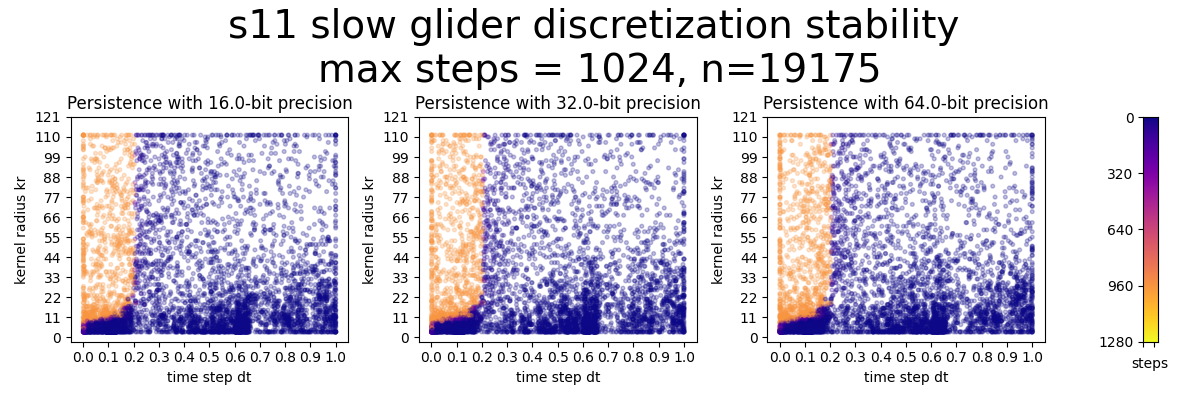}
    \caption{Stability persistence for the s11 slow glider}
\label{fig:s11_map}
\end{figure}

\begin{figure}
\center
  \includegraphics[width=1.0\textwidth]{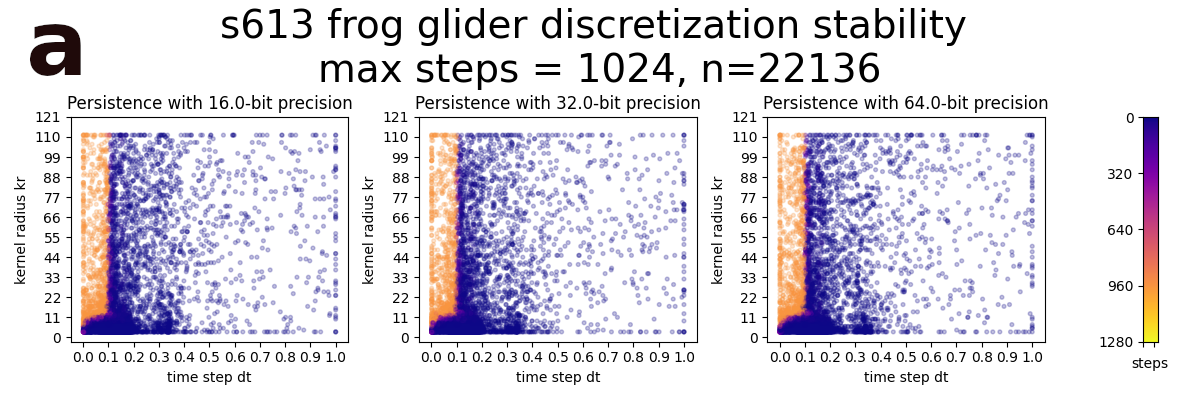}
  \includegraphics[width=1.0\textwidth]{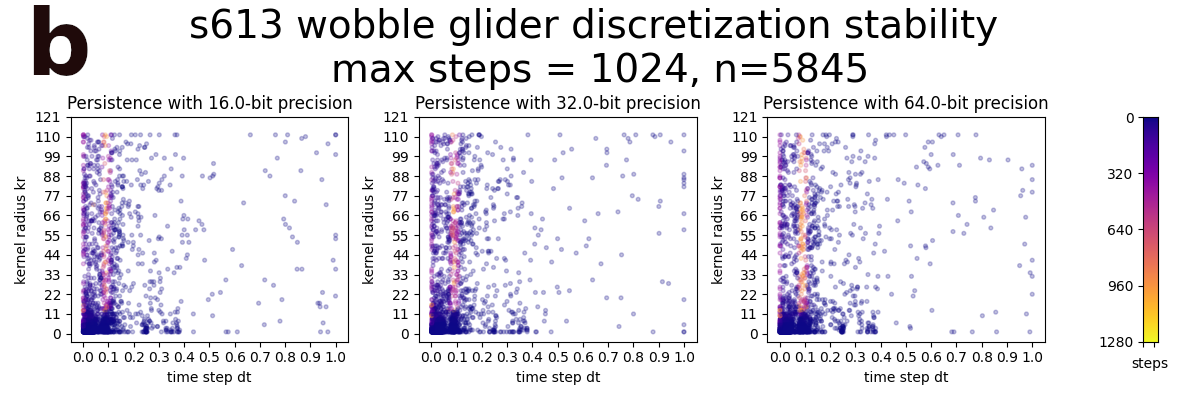}
    \caption{Stability persistence for the s613 frog and s613 wobble gliders}
\label{fig:s613_map}
\end{figure}

\begin{figure}
\center
  \includegraphics[width=1.0\textwidth]{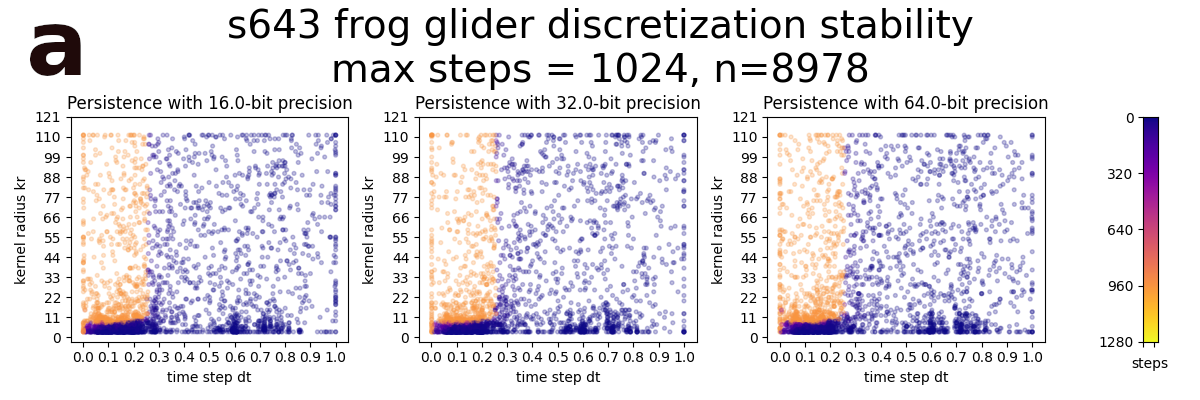}
  \includegraphics[width=1.0\textwidth]{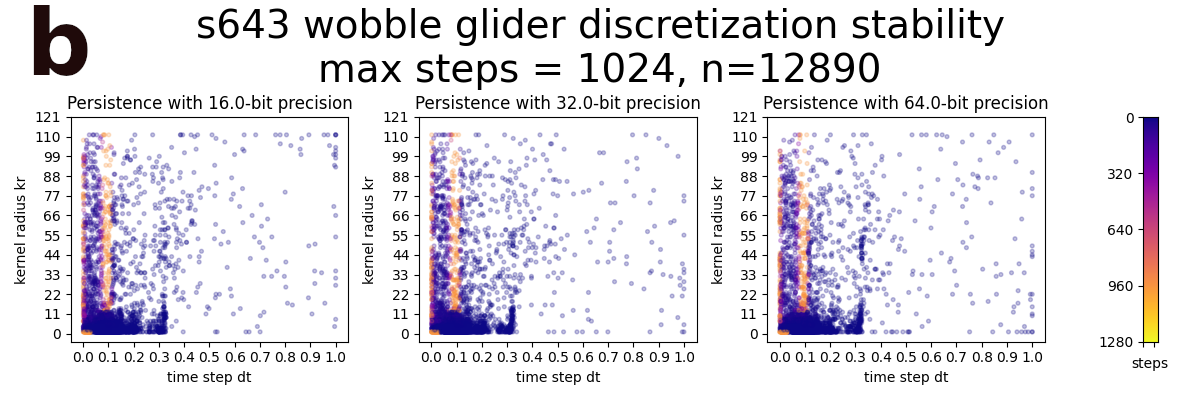}
    \caption{Stability persistence for the s643 frog and s613 wobble gliders}
\label{fig:s643_map}
\end{figure}

The wobble gliders were only semi-stable under the discretization parameters used in their discovery ($\Delta t = 0.1$, $k_r = 31$), and persist for only a few hundred to a few thousand update steps under those conditions. Discretization experiments revealed more stable regions for both wobble gliders, at a step size $\Delta t$ of slightly less than 0.1, but the gliders persist only in a narrow band of $\Delta t$ values and are lost at finer resolution. This is in contrast to the frog gliders in the same rule sets. Those appear to meet criteria for Platonic pattern-rule systems, as does the s11 slow glider. Scatter plots for these pattern-rule pairs are shown in Figures \ref{fig:s11_map} (s11 slow glider), \ref{fig:s643_map}a and \ref{fig:s643_map}b (s643 frog and wobble glider), and \ref{fig:s613_map}a and \ref{fig:s613_map}b (s613 frog and wobble glider). The wobble gliders in s613 and s643 exhibit non-Platonic persistence with respect to numerical precision (Table \ref{tab:dtype}). 


\subsection{Neural Cellular Automata Gliders}

Reflecting their development, neural CA gliders resemble those found in other continuous CA systems: Neurorbium and Neurosynorbium resemble Lenia's {\itshape Orbium} (Figures \ref{fig:glider_collage}l and \ref{fig:glider_collage}n, respectively), the Neurosingle glider (Figure \ref{fig:glider_collage}m) resembles {\itshape Scutium gravidus} and SmoothLife gliders, and the Neurowobble glider is similar to the wobble gliders in {\itshape Hydrogeminium natans} in Lenia, and s643 and s613 in the Glaberish framework.

\begin{figure}
\center
  \includegraphics[width=1.0\textwidth]{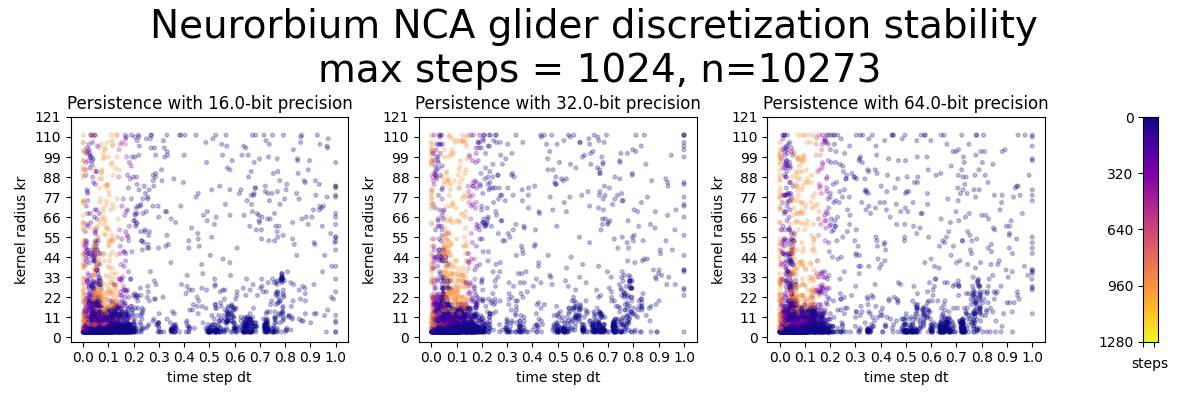}
    \caption{Persistence stability for the Neurorbium glider.}
\label{fig:neurorbium_map}
\end{figure}

\begin{figure}
\center
  \includegraphics[width=1.0\textwidth]{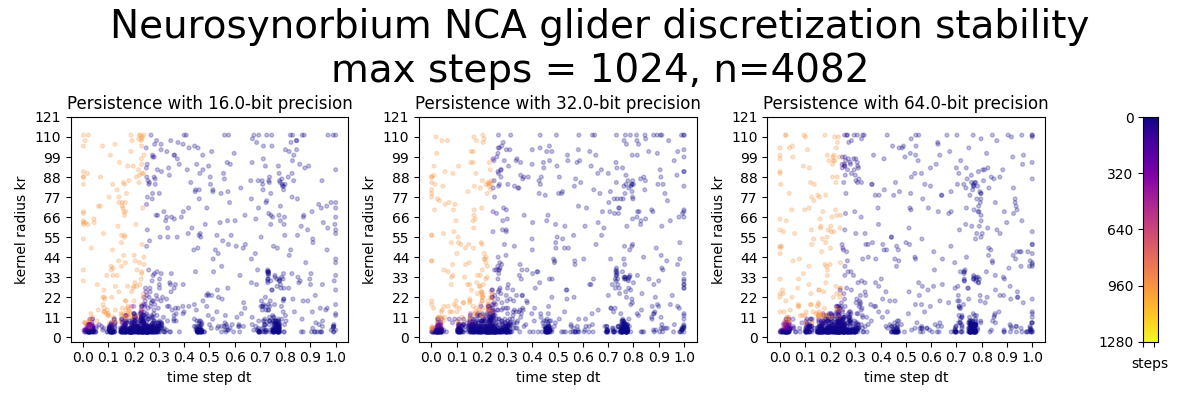}
    \caption{Persistence stability for the Neurosynorbium glider.}
\label{fig:neurosynorbium_map}
\end{figure}

While Neurorbium and Neurosynorbium look similar to the Platonic {\itshape Orbium} Lenia glider, only the latter seems to meet criteria for a Platonic pattern-rule pair (Neurosynorbium, see persistence scatter plot \ref{fig:neurosynorbium_map}). Neurorbium is non-Platonic, losing persistence outside the narrow band of stability at moderate to large kernel sizes and step size $\Delta t$ around 0.1 (Figure \ref{fig:neurorbium_map}). Neurorbium also tends to be stable at very low step sizes around 0.01. In some areas of discretization space, Neurorbium persistence has a complicated, sometimes non-Platonic, dependence on numerical precision (Table \ref{tab:dtype}).

\begin{figure}
\center
  \includegraphics[width=1.0\textwidth]{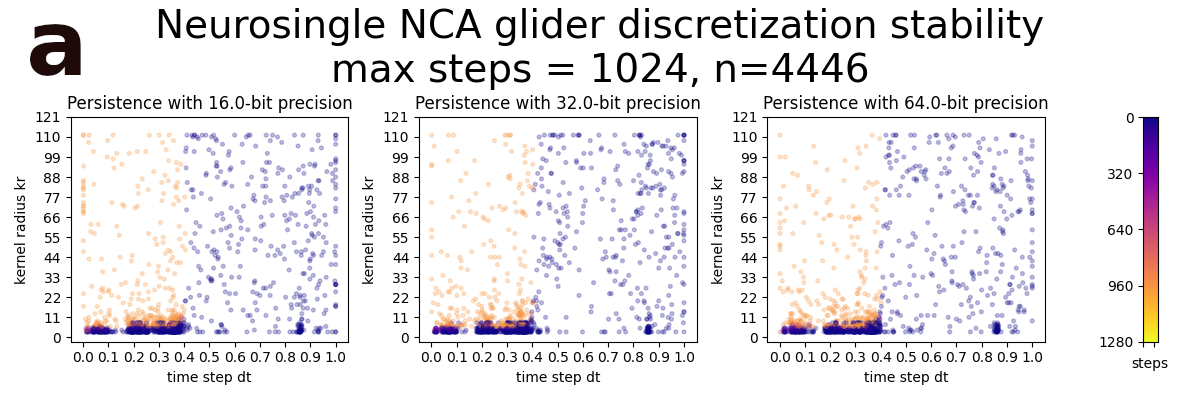}
  \includegraphics[width=1.0\textwidth]{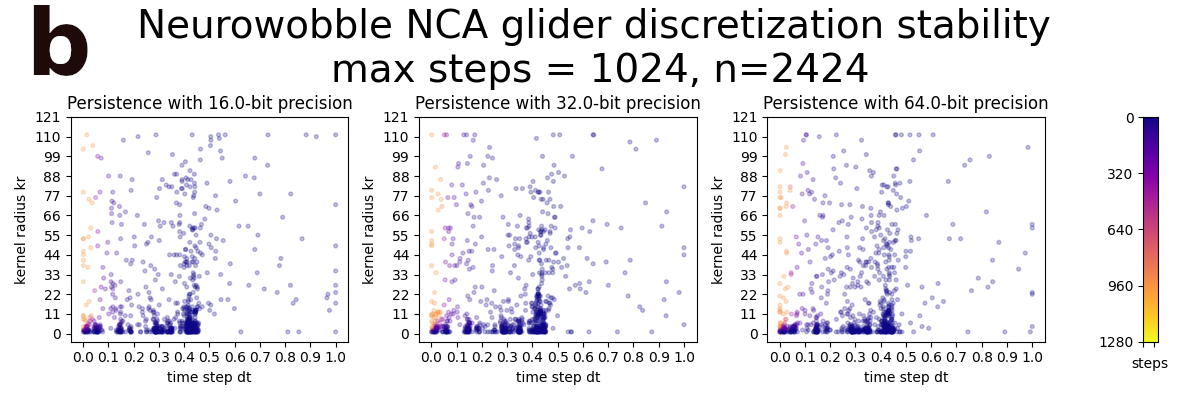}
    \caption{Persistence stability for the Neurosingle and Neurowobble glider, in the neuroscutium\_valvatus NCA.}
\label{fig:neuroscutium_map}
\end{figure}

Unlike the wobble gliders in {\itshape H. natans}, s613, and s643, the Neurowobble glider seems to meet the criteria for a Platonic pattern-rule pair and is only stable at low $\Delta t$ (Figure \ref{fig:neuroscutium_map}b). The Neurosingle glider has a much broader range of support up to $\Delta t$ around 0.4, but also appears Platonic (Figure \ref{fig:neuroscutium_map}a).

\subsection{Preliminary Results in Modified U-Skate World}

The results for glider stability in U-Skate World are not directly comparable to results in continuous and neural CA. Namely, while using the Laplacian of Gaussian function (\ref{eqn:laplacian_of_gaussian}) to approximate the Laplacian $\nabla^2$ in the first terms of Equations \ref{eqn:uskate_u} and \ref{eqn:uskate_v} facilitates scaling kernels to arbitrary size, in practice (in my implementation) only the default 3x3 kernel supports stable gliders. Scaling up the kernel leads to rapid pattern dissolution.

Instead, I followed \citep{munafo2014} and use a $\frac{1}{\Delta x^2}$ coefficient to define spatial resolution. Unlike the CA used in this study, $\frac{1}{\Delta x^2}$ can be thought of as a modification of the diffusion constants $D_u$ and $D_v$. 

In a linear grid search of 20 values from $\Delta x = 6.061*10^{-5}$ to $5.703 * 10^{-3}$ and $\Delta t=0.04$ to $1.0$, the 3 U-Skate gliders considered in this work have a similar range of support, from $2.733*10^{-3}$ to $3.624*10^{-3}$ inclusive. The simple `U-Skate' glider also was stable at $\Delta x=3.921*10^{-3}$. Step size $\Delta t$ does not appear to disrupt the self-organization of these gliders, over the range of values tested here. 

\begin{figure}
\center
  \includegraphics[width=0.8\textwidth]{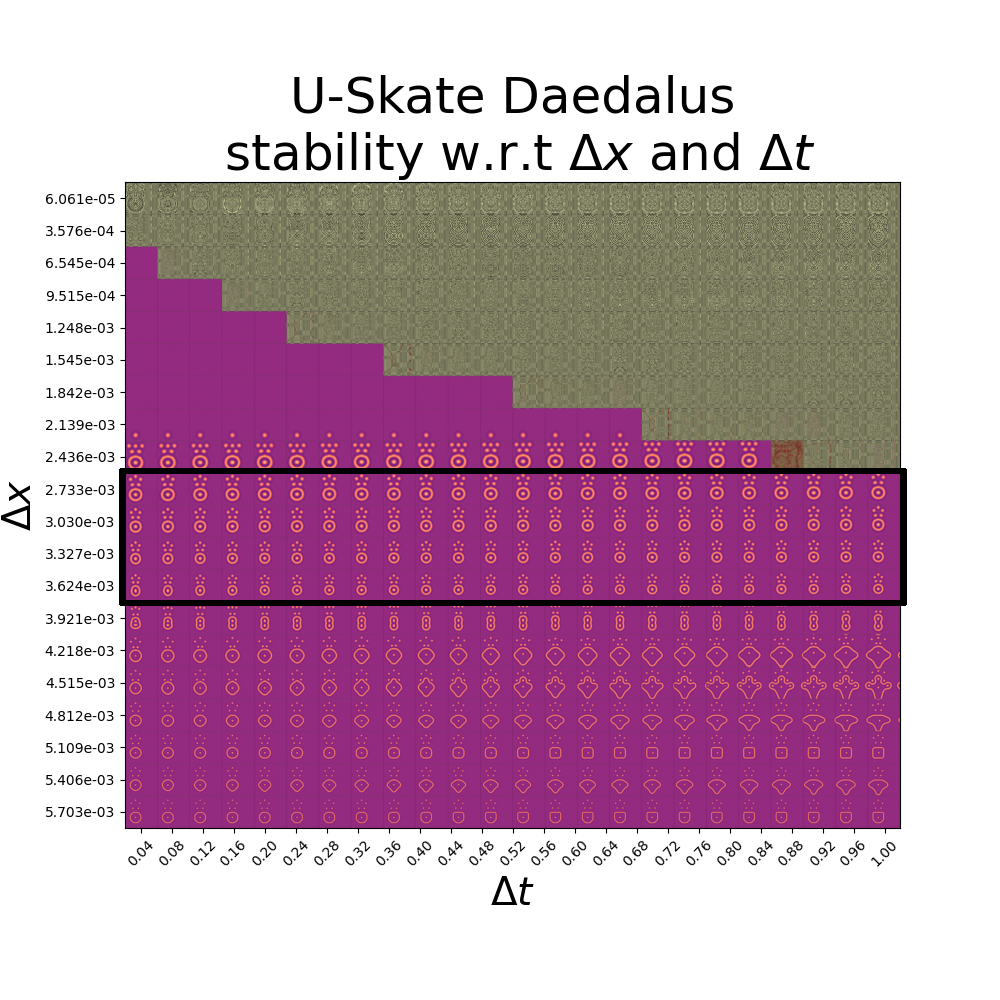}
    \caption{A map of final states for a grid search in $\Delta x$ and $\Delta t$ for the Daedalus glider pattern in modified U-Skate world. The glider appears stable at 262,144 steps for $\Delta x = 2.733e-3$, $\Delta x = 3.030e-3$, $\Delta x = 3.327e-3$, and $\Delta x = 3.624*10^{-3}$.}
\label{fig:daedalus_stability}
\end{figure}

\begin{figure}
\center
  \includegraphics[width=0.8\textwidth]{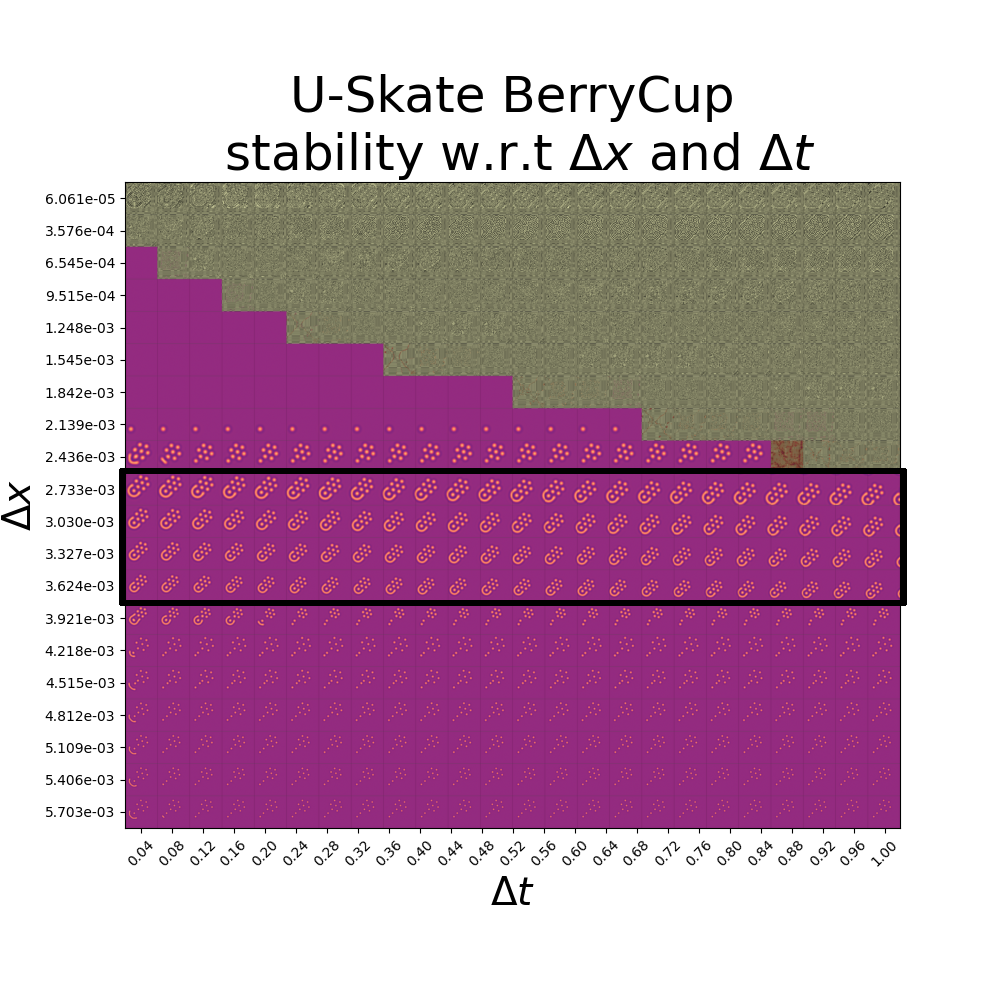}
    \caption{A map of final states for a grid search in $\Delta x$ and $\Delta t$ for the novel BerryCup glider pattern in modified U-Skate world. The glider appears stable at 262,144 steps for $\Delta x = 2.733*10^{-3}$, $\Delta x = 3.030*10^{-3}$, $\Delta x = 3.327*10^{-3}$, and $\Delta x = 3.624*10^{-3}$.}
\label{fig:berrycup_stability}
\end{figure}

\begin{figure}
\center
  \includegraphics[width=0.8\textwidth]{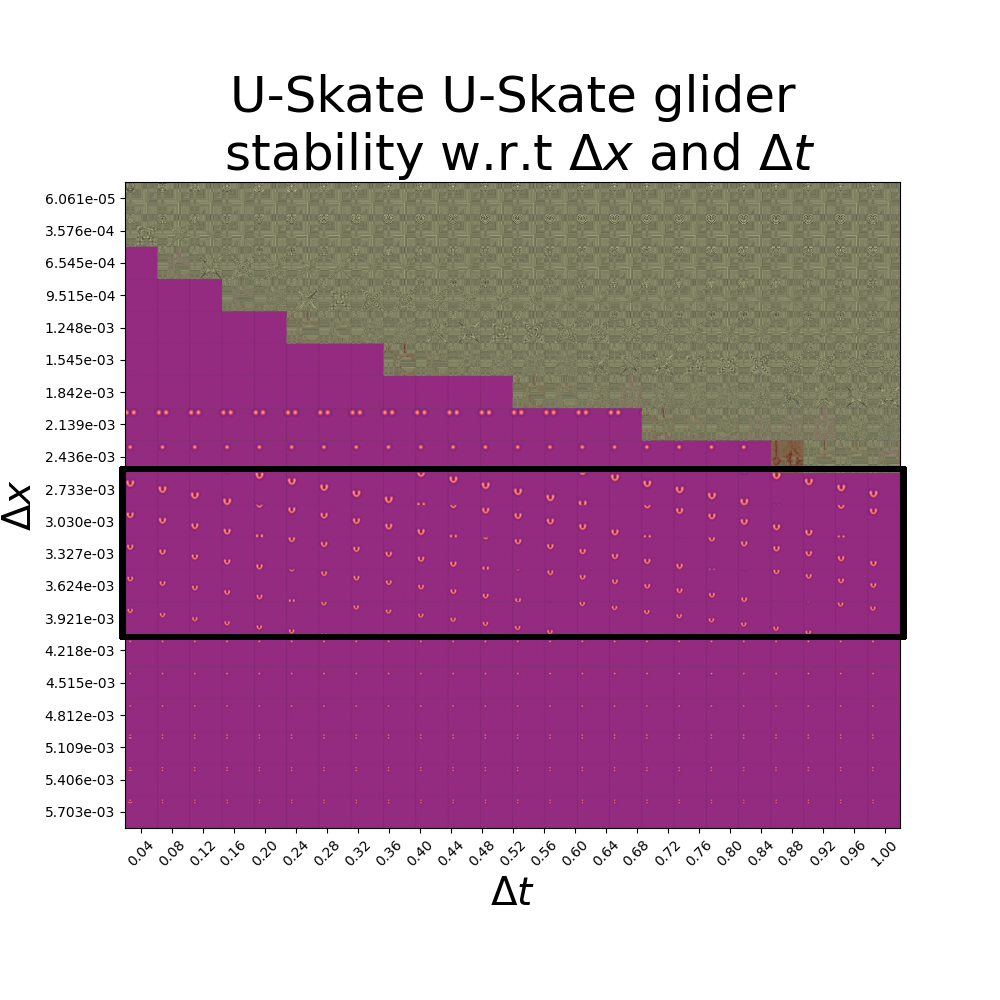}
    \caption{A map of final states for a grid search in $\Delta x$ and $\Delta t$ for the eponymous U-Skate glider pattern in modified U-Skate world. The glider appears stable at 262,144 steps for $\Delta x = 2.733*10^{-3}$, $\Delta x = 3.030*10^{-3}$, $\Delta x = 3.327*10^{-3}$, $\Delta x = 3.624*10^{-3}$, and $\Delta x= 3.921*10^{-3}$.}
\label{fig:uskate_stability}
\end{figure}

Figures \ref{fig:daedalus_stability} and \ref{fig:berrycup_stability} show a map of final grid states for the Daedalus and BerryCup gliders, respectively, after 262,144 steps. End states with a stable glider (as judged by the author) are boxed with a black line. The map of end states for the U-Skate glider is shown in Figure \ref{fig:uskate_stability}, and exhibited glider stability through an additional row ($\Delta x=3.921*10^{-3}$) as well as the same range of stable parameters as the other two gliders. 

The U-Skate gliders in these experiments exhibited persistence that can tentatively be described as non-Platonic, but only with respect to the spatial resolution parameter $\Delta x$ and not $\Delta t$ in the range tested here. As mentioned earlier, $\Delta x$ modifies the diffusion term in Equations \ref{eqn:uskate_u} and \ref{eqn:uskate_v}. The discussion section discusses conflation of rates and discretization as a possible mechanism for non-Platonic glidersd. 

\subsection{Preliminary Results in Adam Optimizer Automaton}

\begin{figure}
\center
  \includegraphics[width=0.6\textwidth]{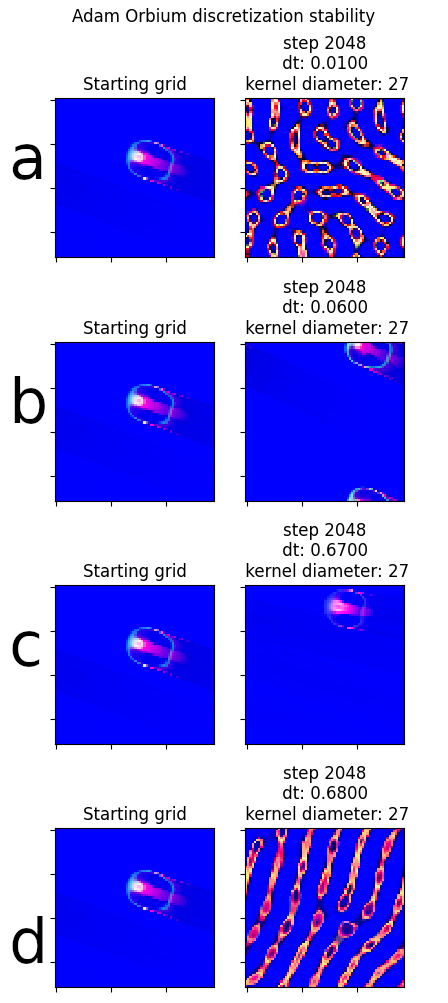}
    \caption{Discretization-dependent dissolution by step 2048 of the `Adorbium' glider in a cellular automaton based on Adam optimizer update rules, at small and large $\Delta t$ step sizes. a) glider is unstable from $\Delta t=0.01$ through $\Delta t=0.05$, with uncontained growth of non-zero cell states. b) glider is stable from $\Delta t=0.06$ through $\Delta t=0.67$ c) glider is again unstable from $\Delta t=0.68$ through $\Delta t=1.0$.}
\label{fig:adorbium}
\end{figure}

The Adorbium pattern-rule pair, unlike the Lenia {\itshape Orbium} glider and rules, is non-Platonic for the parameters investigated in this work. For values of $\Delta t$ of 0.01 to 0.05 with a kernel radius of 13 the glider is unstable; it appears stable after 2048 CA steps for $\Delta t$ values from 0.06 to 0.67, inclusive, and is then unstable for larger $\Delta t$ up to a value of 1.0 (Figure \ref{fig:adorbium}). Exploratory experiments with different grid sizes can vary slightly (in the value of $\Delta t$ that leads to unstable simulations), but qualitatively follow the pattern of instability at low $\Delta t$, stability for a middle range, and instability again for values that are too large.

\begin{figure}
\center
  \includegraphics[width=1.0\textwidth]{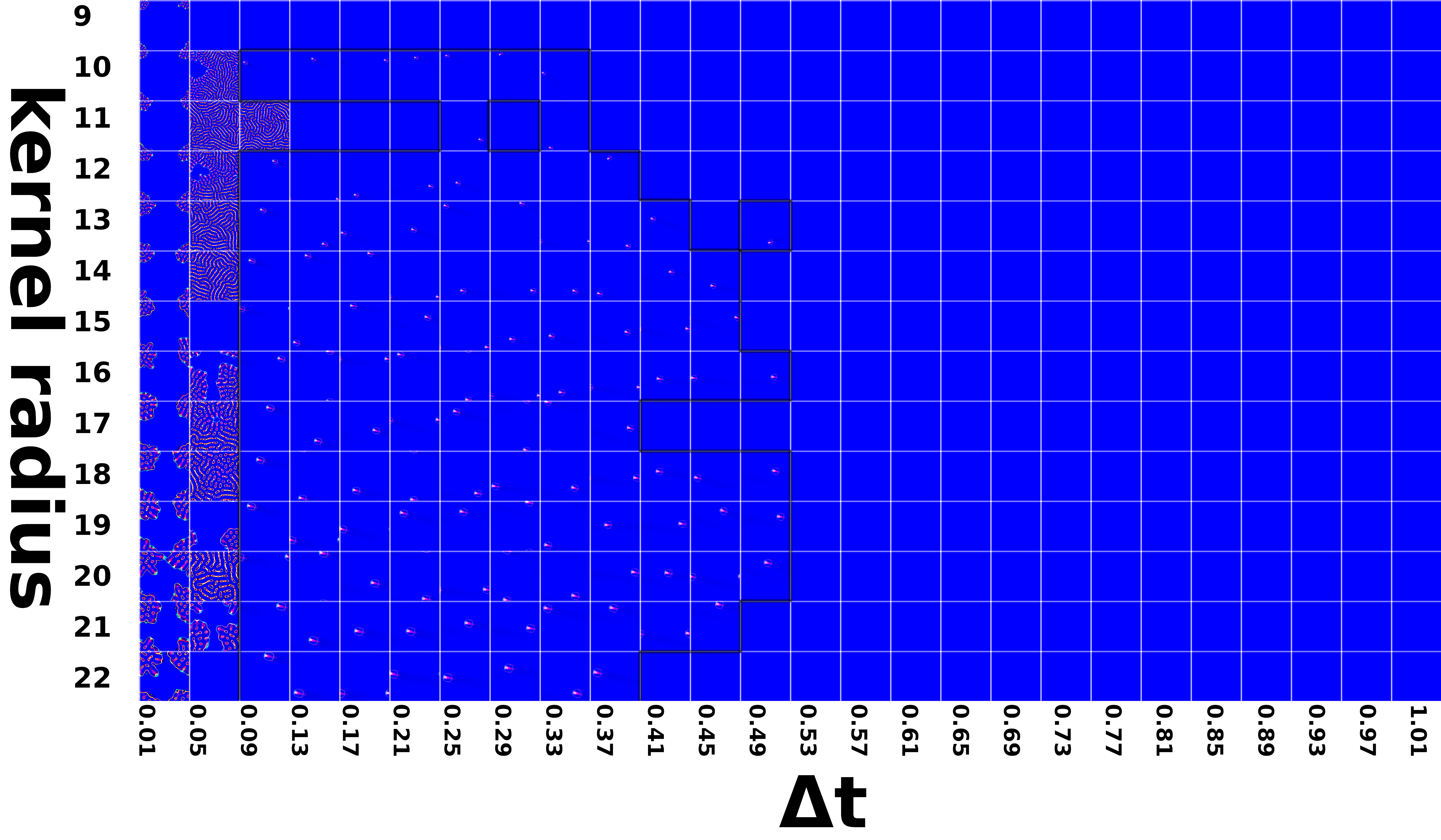}
    \caption{Adorbium grid states at 1024 steps provide an approximate map of glider stability at different values for kernel radius and step size $\Delta t$. Separate grid instances are bordered by lightened pixels, and these boundaries have been darkened by the author to indicate putative stability boundaries, where the glider persisted to step 1024. A full resolution version is included in the supporting resources.} 
\label{fig:adorbium_map}
\end{figure}

In a grid search of kernel radii from 9 to 22 and $\Delta t$ from 0.01 to 1.01 (increments of 0.04) on a common grid size of 180 pixels, Adorbium is again unstable at low $\Delta t$ as well as high $\Delta t$. There appears to be some interdependence of the effects of $\Delta t$ and kernel radius on glider stability, visible in the final grid states in the persistence stability map in Figure \ref{fig:adorbium_map}.


\section{Discussion}

Many glider-rule pairs behave exactly as expected: the glider remains stable as discretization is decreased, in keeping with intuitive expectations associated with simulating physics by numerical methods. About half of glider-rule pairs in my experiments exhibit a more complicated dependence and gliders may become unstable at discretization that is too fine as well as too coarse. In fact, the interaction of different discretization parameters can be synergistic: stability support for a given discretization parameter sometimes depend on other discretization parameters in a non-linear way. 

Systems that support a glider that remains stable for increasingly fine discretization all do so in the same way: given a point in discretization parameter space that is {\it not} stable, increasing coarseness in any discretization parameter never leads to a stable setting. These systems behave as if they are approaching an underlying, abstract ideal as discretization tends to 0 and the system approaches continuity. 

To reflect the relationship of simulation to mathematical idealism, I call these self-organizing systems `Platonic', as their behavior does not conflict with the notion of Platonic idealism. When working with self-organizing pattern-rule combinations that are Platonic, coarse simulations lead to failure but finer simulations do not. 

The situation is markedly different for non-Platonic systems. In non-Platonic systems, the glider is apparently a consequence of simulation artifacts, at least in as much as it depends on a certain level of coarseness to persist. In other words, non-Platonic self-organizing systems are not approximations of an ideal. Their continued existence in effect depends on what would otherwise be considered systematic error. 

Of the 3 hypotheses this project set out to address, 2 of them are not challenged by the results described above. The 2 un-falsified hypotheses predicted that the phenomenon of non-Platonic self-organizing systems would extend to discretization parameters other than step size, and that the effect could be found in multiple complex systems, not just continuous CA. I found non-Platonic self-organization in a wide variety of cellular automata frameworks, including neural cellular automata and a CA based on the adaptive moment update, adam, widely used for training neural networks. In the Gray-Scott artificial chemistry system, the Daedalus, U-Skate, and BerryCup gliders of U-Skate world demonstrate a putative non-Platonism in spatial resolution, though the effect of the spatial parameter $\Delta x$ is conflated with diffusion constants in this system.

The third hypothesis predicted that glider stability would reflect the discretization characteristics of its discovery, {\itshape e.g.} that stability of a glider would be clustered around the discretization parameters where it was evolved, engineered, or discovered. This hypothesis proved to be wrong. In fact, wobble gliders in s643 and s613 are more stable at a smaller step size than the $\Delta t =0.1$ they were evolved with. 

Many combinations of glider patterns and complex systems supporting them adhere to the traditional expectation that greater accuracy can be had with finer discretization, therefore the essential role of discretization is not a general rule. But I did find examples of non-Platonic combinations in every type of system investigated in this work. There does not appear to be an easy way to predict {\itshape a priori} whether a given self-organizing pattern in a given system will be Platonic or non-Platonic.

The relationship between discretization and glider stability is not dependent on either the rules of a system nor the glider pattern in isolation. Continuous CA {\itshape Hydrogeminium natans}, s613, and s643 are all Platonic in the context of at least one glider, and non-Platonic in simulating another. The {\itshape Orbium}, Neurorbium, Neurosynorbium, and Adorbium gliders all look alike, but Adorbium and Neurorbium are non-Platonic while the rest are Platonic. The same can be said for other groupings of similar glider patterns: two of three of the similar {\itshape Scutium gravidus}, SmoothLife, and Neurosingle gliders are non-Platonc, as are four of the five wide/wobble gliders supported in {\itshape H. natans}, {\itshape Triscutium solidus}, s613, s643, and neuroscutium\_valvatus rule sets. 

Perhaps the most interesting non-Platonic glider-rule pair found in this work is the SmoothLife glider. The support region for this glider demonstrates a marked synergy between spatial and temporal resolution (Figure \ref{fig:smoothlife_map}). In most instances of non-Platonism described in this work spatial and temporal discretization had little effect on each other, and support regions are bounded in roughly rectangular areas (though there may be multiple regions of support). The SmoothLife glider in contrast has a distinctly curved support region. 

The diversity of scenarios where non-Platonic self-organization can be found suggests to this author that the cause is likely a systems effect that depends on multiple interacting mechanisms, but one plausible contributing factor is the conflation of rates described by update equations and the step size. To explore the concept of rate conflation, we can consider discretization effects in two well known systems, the Lorenz system and the (continuous version of) the logistic map \citep{lorenz1963, may1976}. These classic examples from complex systems science and chaos theory have been meticulously studied over the years, and if we can understand non-Platonic gliders in similar terms the tools developed for analyzing them may prove useful.  

\begin{figure}
\center
  \includegraphics[width=0.8\textwidth]{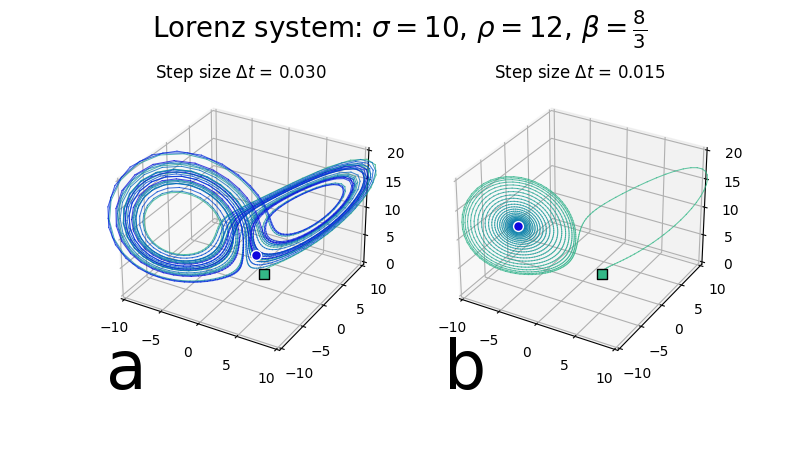}
    \caption{The Lorenz system with $\sigma = 10$, $\rho = 12$, and $\beta = \frac{8}{3}$ and initial conditions $(x_0, y_0, z_0) = (2, 1, 1)$ settles to a quiescent steady state at $(x, y, z) \approx (-5.416, -5.416, 11.0)$ when simulated with a sufficiently fine step size of 0.015. Simulating the system with the same parameters and initial conditions, but a step size of $\Delta t = 0.03$ yields a chaotic trajectory with a strange attractor.}
\label{fig:lorenz}
\end{figure}

Perhaps the most mundane explanation for non-Platonic behavior is that the differential equations are wrong, and using coarse discretization artificially approximates a more suitable set of equations, with different rates of change, for a given pattern. The Lorenz system\footnote{Characterized by equations $\frac{dx}{dt} = \sigma(y-x)$, $\frac{dy}{dt} = x(\rho-z)-y$, $\frac{dz}{dt} = xy - \beta z$ } is a set of three differential equations that, given the right parameters and initial conditions, generates chaotic trajectories with a strange attractor. Simulated at high resolution, Lorenz system parameters $\sigma = 10$, $\rho=12$, and $\beta=\frac{8}{3}$ with initial conditions $x=2$, $y=1$, $z=1$, the Lorenz system has a `boring' trajectory: the system eventually reaches a quiescent steady state at $x=-5.416$, $y=-5.416$ and $z=11.00$ (Figure\ref{fig:lorenz}a). Retaining the same parameters and initial conditions but doubling the step size from $\Delta t = 0.015$ to $\Delta t = 0.03$ yields a chaotic trajectory that superficially resembles the type of behavior that chaos mathematicians might find more interesting (Figure \ref{fig:lorenz}b).

Conflation of the rate of change defining these systems with the step size used to simulate them is one plausible route to non-Platonic behavior, as we can see hints of in the Lorenz system example. In pattern-rule pairs where this occurs, it could explain some of the dependence of non-Platonic behavior on spatial resolution, as the denominator of speed. If it occurs, this could yield a signature in the range of discretization support for a given pattern and rule, for example in a relationship between spatial and temporal discretization that satisfies the Courant-Friedrichs-Lewy (CFL) condition or something similar \citep{courant1967}\footnote{The Courant number is the ratio of the product of the step size with the speed of a traveling phenomenon and the size of a grid point, {\itshape i.e.} for one dimension $\frac{v \Delta t}{\Delta x}$, where v is speed.}. I did not see clear evidence for this from my experiments, but the interaction may be a more complex, non-linear one. Most support regions for non-Platonic gliders are roughly rectangular, but SmoothLife does show pronounced interaction between $\Delta t$ and kernel radius $k_r$ that could be consistent with self-organizing persistence from artifacts arising from a failure to satisfy something like the CFL condition. 

Another piece must be added to explain the sometimes observable dependence of non-Platonic persistence with respect to numerical precision, however. To gain insight into this aspect, we will briefly visit another classic example of complexity and chaos theory in the differential equation of the logistic. 

\begin{figure}
\center
  \includegraphics[width=0.8\textwidth]{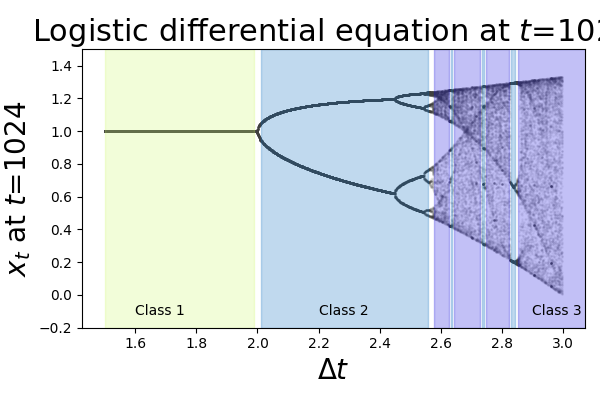}
    \caption{Bifurcation diagram of the logistic differential equation. Areas that tend to a quiescent steady state, oscillations, or choas are approximately marked as class 1, class 2, and class 3 (analogous to CA classifications), respectively.}
\label{fig:logistic_deq}
\end{figure}

The logistic function is a solution to the differential equation $\frac{\partial f(t)}{\partial t} = f(t)(1-f(t))$ with the boundary condition that $f(0) = \frac{1}{2}$. Likely more familiar is the discrete version, known as the logistic map, $f(t+1) = rf(t)(1-f(t))$, where rate $r$ determines by what magnitude $f(t+1)$ will differ from $f(t)$. If we simulate the logistic differential equation across a range of step sizes $\Delta t$ (conflated with an implicit rate $r=1$), we get a  bifurcation diagram very similar to that of the discrete logistic map. This bifurcation diagram demonstrates regions of quiescent equilibrium, periodic oscillations, and chaos (Figure \ref{fig:logistic_deq}). A common classification scheme categorizes CA based on a tendency to reach a quiescent stable state (class 1), static and oscillatory patterns (class 2), and chaos (class 3), and a fourth class characterized by regions of order and chaos \citep{wolfram1984}. Many class 4 CA, including Conway's Game of Life and elementary CA Rule 110, are computationally universal \citep{berlekamp2004, cook2004}.

The analogy between quiescent, periodic, and chaotic regions in the bifurcation diagram in Figure \ref{fig:logistic_deq} and class 1, 2, and 3 CA may be a useful clue. CA considered most interesting fall under class 4, exhibiting chaos and order in complex sequence across space and time. Perhaps some continuous CA (and other complex systems) are intrinsically balanced between chaos and order, and the spatial distribution of patterns is sufficient to give rise to distinct regions of order and chaos. Other continuous CA and complex systems may be balanced in a region of rule space analogous to the bifurcation nodes in Figure \ref{fig:logistic_deq} between order and chaos, but need a combination of discretization errors (possibly floating point rounding errors) and particular patterns to `tip the balance' between a diverse repertoire of chaotic and ordered behavior. If this quasi-stable explanation for non-Platonic behavior is true, it should be possible to engineer simple systems like the logistic differential equation to both `knock out' and `rescue' non-Platonic behavior\footnote{This terminology is adopted from genetics, and complements Feynman's famous quote about the route to understanding by building. Removing a gene that destroys a phenotype might indicate it is responsible for a trait, but for greater confidence it should be possible to restore the trait by re-introducing the gene.}.



\label{sec:discussion}




\section{Conclusion}

In ``Concept of Mind'', philosopher Gilbert Ryle introduces the phrase ``Ghost in the Machine'' as a criticism-via-absurdity of Cartesian dualism \citep{ryle1949}. In Ryle's view, the idea that the mind (or soul) is a separate thing causally disconnected from the body and the material world is as ridiculous as a ghost in a machine. In contemporary usage (especially fiction) the term has come to mean the opposite of its original intention of {\itshape reductio ad absurdum}: ``Ghost in the Machine'' is more likely to refer to the surprising emergence of intelligence, agency, and/or self-organization in a sufficiently complex system, without being explicitly programmed or designed. This is a potent metaphor for the emergence of self-organization in complex systems like those studied here, and the results of this work are fully in agreement with the inextricable relationship between ``Machines'', complex systems' rules and their implementation artifacts, and ``Ghosts'', the forms that arise and persist therein.

The state-plus-update pattern exemplified in the Euler method is widely applicable in numerical simulations. It is a common component of physics simulations (simple ones, at least) of the physics of our own universe as well as the artificial universes considered in this article. It is also evident in the residual connections of neural networks, the incremental updates of optimization procedures like adam optimization or stochastic gradient descent, and from a high-level view reflects the feedback process in many control systems.

It remains to be seen if non-Platonic self-organization occurs in more sophisticated numerical methods, the kind that might be trusted to simulate and predict the future. But the effect of discretization on self-organization goes beyond the risk of self-organizing entities like a weather pattern or climate tipping point arising in simulation from a poor choice of discretization parameters. 

Discretization also seems to play an intrinsic part in human experience. The discovery of period-doubling bifurcation in responses to flickering illumination, remarkably similar to Figure \ref{fig:logistic_deq}, in the visual systems of humans and salamanders \citep{crevier1998}, aliasing effects in human vision under continuous lighting that occur in some situations \citep{purves1996}, and other `freeze-frame' illusions reported under the influence of drugs or other neurological conditions \citep{dubois2011}, all point to an ambiguous but interesting role for discretization in human consciousness and perception.


\label{sec:conclusion}

\footnotesize
\bibliographystyle{apalike}
\bibliography{references} 

\begin{thebibliography}{}

\bibitem [\protect \citeauthoryear {%
Bergstra%
\ \BBA {} Bengio%
}{%
Bergstra%
\ \BBA {} Bengio%
}{%
{\protect \APACyear {2012}}%
}]{%
bergstra2012}
\APACinsertmetastar {%
bergstra2012}%
\begin{APACrefauthors}%
Bergstra, J.%
\BCBT {}\ \BBA {} Bengio, Y.%
\end{APACrefauthors}%
\unskip\
\newblock
\APACrefYearMonthDay{2012}{}{}.
\newblock
{\BBOQ}\APACrefatitle {Random Search for Hyper-Parameter Optimization} {Random search for hyper-parameter optimization}.{\BBCQ}
\newblock
\APACjournalVolNumPages{J. Mach. Learn. Res.}{13}{}{281-305}.
\PrintBackRefs{\CurrentBib}

\bibitem [\protect \citeauthoryear {%
Berlekamp%
, Conway%
\BCBL {}\ \BBA {} Guy%
}{%
Berlekamp%
\ \protect \BOthers {.}}{%
{\protect \APACyear {2004}}%
}]{%
berlekamp2004}
\APACinsertmetastar {%
berlekamp2004}%
\begin{APACrefauthors}%
Berlekamp, E\BPBI R.%
, Conway, J\BPBI H.%
\BCBL {}\ \BBA {} Guy, R\BPBI K.%
\end{APACrefauthors}%
\unskip\
\newblock
\APACrefYear{2004}.
\newblock
\APACrefbtitle {Winning Ways for Your Mathematical Plays Volume 4. Second Edition} {Winning ways for your mathematical plays volume 4. second edition}.
\newblock
\APACaddressPublisher{}{A K Peters, Wellesly, Massachusetts}.
\PrintBackRefs{\CurrentBib}

\bibitem [\protect \citeauthoryear {%
Chan%
}{%
Chan%
}{%
{\protect \APACyear {2019}}%
}]{%
chan2019}
\APACinsertmetastar {%
chan2019}%
\begin{APACrefauthors}%
Chan, B\BPBI W\BHBI C.%
\end{APACrefauthors}%
\unskip\
\newblock
\APACrefYearMonthDay{2019}{}{}.
\newblock
{\BBOQ}\APACrefatitle {{Lenia} - Biology of Artificial Life} {{Lenia} - biology of artificial life}.{\BBCQ}
\newblock
\APACjournalVolNumPages{Complex Systems}{28}{3}{251--286}.
\newblock
\begin{APACrefURL} [{2021-11-23}]\url{http://arxiv.org/abs/1812.05433} \end{APACrefURL}
\newblock
\begin{APACrefDOI} \doi{10.25088/ComplexSystems.28.3.251} \end{APACrefDOI}
\PrintBackRefs{\CurrentBib}

\bibitem [\protect \citeauthoryear {%
Cook%
\ \protect \BOthers {.}}{%
Cook%
\ \protect \BOthers {.}}{%
{\protect \APACyear {2004}}%
}]{%
cook2004}
\APACinsertmetastar {%
cook2004}%
\begin{APACrefauthors}%
Cook, M.%
\BCBT {}\ \BOthersPeriod {.}
\end{APACrefauthors}%
\unskip\
\newblock
\APACrefYearMonthDay{2004}{}{}.
\newblock
{\BBOQ}\APACrefatitle {Universality in elementary cellular automata} {Universality in elementary cellular automata}.{\BBCQ}
\newblock
\APACjournalVolNumPages{Complex systems}{15}{1}{1--40}.
\PrintBackRefs{\CurrentBib}

\bibitem [\protect \citeauthoryear {%
Courant%
, Friedrichs%
\BCBL {}\ \BBA {} Lewy%
}{%
Courant%
\ \protect \BOthers {.}}{%
{\protect \APACyear {1967}}%
}]{%
courant1967}
\APACinsertmetastar {%
courant1967}%
\begin{APACrefauthors}%
Courant, R.%
, Friedrichs, K.%
\BCBL {}\ \BBA {} Lewy, H.%
\end{APACrefauthors}%
\unskip\
\newblock
\APACrefYearMonthDay{1967}{}{}.
\newblock
{\BBOQ}\APACrefatitle {On the partial difference equations of mathematical physics} {On the partial difference equations of mathematical physics}.{\BBCQ}
\newblock
\APACjournalVolNumPages{IBM journal of Research and Development}{11}{2}{215--234}.
\PrintBackRefs{\CurrentBib}

\bibitem [\protect \citeauthoryear {%
Crevier%
\ \BBA {} Meister%
}{%
Crevier%
\ \BBA {} Meister%
}{%
{\protect \APACyear {1998}}%
}]{%
crevier1998}
\APACinsertmetastar {%
crevier1998}%
\begin{APACrefauthors}%
Crevier, D\BPBI W.%
\BCBT {}\ \BBA {} Meister, M.%
\end{APACrefauthors}%
\unskip\
\newblock
\APACrefYearMonthDay{1998}{}{}.
\newblock
{\BBOQ}\APACrefatitle {Synchronous period-doubling in flicker vision of salamander and man} {Synchronous period-doubling in flicker vision of salamander and man}.{\BBCQ}
\newblock
\APACjournalVolNumPages{Journal of neurophysiology}{79}{4}{1869--1878}.
\PrintBackRefs{\CurrentBib}

\bibitem [\protect \citeauthoryear {%
Davis%
\ \BBA {} Bongard%
}{%
Davis%
\ \BBA {} Bongard%
}{%
{\protect \APACyear {2022}}%
{\protect \APACexlab {{\protect \BCnt {1}}}}}]{%
davis2022a}
\APACinsertmetastar {%
davis2022a}%
\begin{APACrefauthors}%
Davis, Q\BPBI T.%
\BCBT {}\ \BBA {} Bongard, J.%
\end{APACrefauthors}%
\unskip\
\newblock
\APACrefYearMonthDay{2022{\protect \BCnt {1}}}{07}{}.
\newblock
{\BBOQ}\APACrefatitle {{Glaberish: Generalizing the Continuously-Valued Lenia Framework to Arbitrary Life-Like Cellular Automata}} {{Glaberish: Generalizing the Continuously-Valued Lenia Framework to Arbitrary Life-Like Cellular Automata}}.{\BBCQ}
\newblock
\APACjournalVolNumPages{ALIFE 2022: The 2022 Conference on Artificial Life}{ALIFE 2022: The 2022 Conference on Artificial Life}{}{}.
\newblock
\begin{APACrefURL} \url{https://doi.org/10.1162/isal\_a\_00530} \end{APACrefURL}
\newblock
\APACrefnote{47}
\newblock
\begin{APACrefDOI} \doi{10.1162/isal_a_00530} \end{APACrefDOI}
\PrintBackRefs{\CurrentBib}

\bibitem [\protect \citeauthoryear {%
Davis%
\ \BBA {} Bongard%
}{%
Davis%
\ \BBA {} Bongard%
}{%
{\protect \APACyear {2022}}%
{\protect \APACexlab {{\protect \BCnt {3}}}}}]{%
davis2022b}
\APACinsertmetastar {%
davis2022b}%
\begin{APACrefauthors}%
Davis, Q\BPBI T.%
\BCBT {}\ \BBA {} Bongard, J.%
\end{APACrefauthors}%
\unskip\
\newblock
\APACrefYearMonthDay{2022{\protect \BCnt {3}}}{07}{}.
\newblock
{\BBOQ}\APACrefatitle {Step Size is a Consequential Parameter in Continuous Cellular Automata} {Step size is a consequential parameter in continuous cellular automata}.{\BBCQ}
\newblock
\APACjournalVolNumPages{ALIFE 2022: The 2022 Conference on Artificial Life}{43}{}{}.
\newblock
\begin{APACrefURL} \url{https://doi.org/10.1162/isal\_a\_00526} \end{APACrefURL}
\newblock
\begin{APACrefDOI} \doi{10.1162/isal_a_00526} \end{APACrefDOI}
\PrintBackRefs{\CurrentBib}

\bibitem [\protect \citeauthoryear {%
Davis%
\ \BBA {} Bongard%
}{%
Davis%
\ \BBA {} Bongard%
}{%
{\protect \APACyear {2022}}%
{\protect \APACexlab {{\protect \BCnt {2}}}}}]{%
davis2022c}
\APACinsertmetastar {%
davis2022c}%
\begin{APACrefauthors}%
Davis, Q\BPBI T.%
\BCBT {}\ \BBA {} Bongard, J\BPBI C.%
\end{APACrefauthors}%
\unskip\
\newblock
\APACrefYearMonthDay{2022{\protect \BCnt {2}}}{}{}.
\newblock
{\BBOQ}\APACrefatitle {Selecting continuous life-like cellular automata for halting unpredictability: evolving for abiogenesis} {Selecting continuous life-like cellular automata for halting unpredictability: evolving for abiogenesis}.{\BBCQ}
\newblock
\APACjournalVolNumPages{Proceedings of the Genetic and Evolutionary Computation Conference Companion}{}{}{}.
\newblock
\begin{APACrefURL} \url{https://doi.org/10.1145/3520304.3529037} \end{APACrefURL}
\PrintBackRefs{\CurrentBib}

\bibitem [\protect \citeauthoryear {%
Dubois%
\ \BBA {} VanRullen%
}{%
Dubois%
\ \BBA {} VanRullen%
}{%
{\protect \APACyear {2011}}%
}]{%
dubois2011}
\APACinsertmetastar {%
dubois2011}%
\begin{APACrefauthors}%
Dubois, J.%
\BCBT {}\ \BBA {} VanRullen, R.%
\end{APACrefauthors}%
\unskip\
\newblock
\APACrefYearMonthDay{2011}{}{}.
\newblock
{\BBOQ}\APACrefatitle {Visual trails: do the doors of perception open periodically?} {Visual trails: do the doors of perception open periodically?}{\BBCQ}
\newblock
\APACjournalVolNumPages{PLoS biology}{9}{5}{e1001056}.
\PrintBackRefs{\CurrentBib}

\bibitem [\protect \citeauthoryear {%
Gardner%
}{%
Gardner%
}{%
{\protect \APACyear {1970}}%
}]{%
gardner1970}
\APACinsertmetastar {%
gardner1970}%
\begin{APACrefauthors}%
Gardner, M.%
\end{APACrefauthors}%
\unskip\
\newblock
\APACrefYearMonthDay{1970}{}{}.
\newblock
{\BBOQ}\APACrefatitle {Mathematical Games: The Fantastic Combinations of {John Conway}'s New Solitaire Game ``{Life}"} {Mathematical games: The fantastic combinations of {John Conway}'s new solitaire game ``{Life}"}.{\BBCQ}
\newblock
\APACjournalVolNumPages{Scientific American}{223}{}{120-123}.
\newblock
\begin{APACrefURL} \url{https://web.stanford.edu/class/sts145/Library/life.pdf} \end{APACrefURL}
\PrintBackRefs{\CurrentBib}

\bibitem [\protect \citeauthoryear {%
Gray%
\ \BBA {} Scott%
}{%
Gray%
\ \BBA {} Scott%
}{%
{\protect \APACyear {1985}}%
}]{%
gray1985}
\APACinsertmetastar {%
gray1985}%
\begin{APACrefauthors}%
Gray, P.%
\BCBT {}\ \BBA {} Scott, S\BPBI K.%
\end{APACrefauthors}%
\unskip\
\newblock
\APACrefYearMonthDay{1985}{}{}.
\newblock
{\BBOQ}\APACrefatitle {Sustained oscillations and other exotic patterns of behavior in isothermal reactions} {Sustained oscillations and other exotic patterns of behavior in isothermal reactions}.{\BBCQ}
\newblock
\APACjournalVolNumPages{The Journal of Physical Chemistry}{89}{}{22-32}.
\PrintBackRefs{\CurrentBib}

\bibitem [\protect \citeauthoryear {%
Kingma%
\ \BBA {} Ba%
}{%
Kingma%
\ \BBA {} Ba%
}{%
{\protect \APACyear {2014}}%
}]{%
kingma2014}
\APACinsertmetastar {%
kingma2014}%
\begin{APACrefauthors}%
Kingma, D\BPBI P.%
\BCBT {}\ \BBA {} Ba, J.%
\end{APACrefauthors}%
\unskip\
\newblock
\APACrefYearMonthDay{2014}{}{}.
\newblock
{\BBOQ}\APACrefatitle {Adam: A Method for Stochastic Optimization} {Adam: A method for stochastic optimization}.{\BBCQ}
\newblock
\APACjournalVolNumPages{CoRR}{abs/1412.6980}{}{}.
\PrintBackRefs{\CurrentBib}

\bibitem [\protect \citeauthoryear {%
Lorenz%
}{%
Lorenz%
}{%
{\protect \APACyear {1963}}%
}]{%
lorenz1963}
\APACinsertmetastar {%
lorenz1963}%
\begin{APACrefauthors}%
Lorenz, E\BPBI N.%
\end{APACrefauthors}%
\unskip\
\newblock
\APACrefYearMonthDay{1963}{}{}.
\newblock
{\BBOQ}\APACrefatitle {Deterministic nonperiodic flow} {Deterministic nonperiodic flow}.{\BBCQ}
\newblock
\APACjournalVolNumPages{Journal of atmospheric sciences}{20}{2}{130--141}.
\PrintBackRefs{\CurrentBib}

\bibitem [\protect \citeauthoryear {%
{MacLennan}%
}{%
{MacLennan}%
}{%
{\protect \APACyear {1990}}%
}]{%
maclennan1990}
\APACinsertmetastar {%
maclennan1990}%
\begin{APACrefauthors}%
{MacLennan}, B\BPBI J.%
\end{APACrefauthors}%
\unskip\
\newblock
\APACrefYearMonthDay{1990}{}{}.
\newblock
{\BBOQ}\APACrefatitle {Continuous Spatial Automata} {Continuous spatial automata}.{\BBCQ}
\newblock

\PrintBackRefs{\CurrentBib}

\bibitem [\protect \citeauthoryear {%
May%
}{%
May%
}{%
{\protect \APACyear {1976}}%
}]{%
may1976}
\APACinsertmetastar {%
may1976}%
\begin{APACrefauthors}%
May, R\BPBI M.%
\end{APACrefauthors}%
\unskip\
\newblock
\APACrefYearMonthDay{1976}{}{}.
\newblock
{\BBOQ}\APACrefatitle {Simple mathematical models with very complicated dynamics} {Simple mathematical models with very complicated dynamics}.{\BBCQ}
\newblock
\APACjournalVolNumPages{Nature}{261}{5560}{459--467}.
\PrintBackRefs{\CurrentBib}

\bibitem [\protect \citeauthoryear {%
Mordvintsev%
, Randazzo%
, Niklasson%
\BCBL {}\ \BBA {} Levin%
}{%
Mordvintsev%
\ \protect \BOthers {.}}{%
{\protect \APACyear {2020}}%
}]{%
mordvintsev2020}
\APACinsertmetastar {%
mordvintsev2020}%
\begin{APACrefauthors}%
Mordvintsev, A.%
, Randazzo, E.%
, Niklasson, E.%
\BCBL {}\ \BBA {} Levin, M.%
\end{APACrefauthors}%
\unskip\
\newblock
\APACrefYearMonthDay{2020}{}{}.
\newblock
{\BBOQ}\APACrefatitle {Growing Neural Cellular Automata} {Growing neural cellular automata}.{\BBCQ}
\newblock
\APACjournalVolNumPages{Distill}{}{}{}.
\newblock
\APACrefnote{https://distill.pub/2020/growing-ca}
\newblock
\begin{APACrefDOI} \doi{10.23915/distill.00023} \end{APACrefDOI}
\PrintBackRefs{\CurrentBib}

\bibitem [\protect \citeauthoryear {%
Munafo%
}{%
Munafo%
}{%
{\protect \APACyear {2014}}%
}]{%
munafo2014}
\APACinsertmetastar {%
munafo2014}%
\begin{APACrefauthors}%
Munafo, R.%
\end{APACrefauthors}%
\unskip\
\newblock
\APACrefYearMonthDay{2014}{}{}.
\newblock
{\BBOQ}\APACrefatitle {Stable localized moving patterns in the 2-D Gray-Scott model} {Stable localized moving patterns in the 2-d gray-scott model}.{\BBCQ}
\newblock
\APACjournalVolNumPages{arXiv: Pattern Formation and Solitons}{}{}{}.
\PrintBackRefs{\CurrentBib}

\bibitem [\protect \citeauthoryear {%
Niklasson%
, Mordvintsev%
, Randazzo%
\BCBL {}\ \BBA {} Levin%
}{%
Niklasson%
\ \protect \BOthers {.}}{%
{\protect \APACyear {2021}}%
}]{%
niklasson2021}
\APACinsertmetastar {%
niklasson2021}%
\begin{APACrefauthors}%
Niklasson, E.%
, Mordvintsev, A.%
, Randazzo, E.%
\BCBL {}\ \BBA {} Levin, M.%
\end{APACrefauthors}%
\unskip\
\newblock
\APACrefYearMonthDay{2021}{}{}.
\newblock
{\BBOQ}\APACrefatitle {Self-Organising Textures} {Self-organising textures}.{\BBCQ}
\newblock
\APACjournalVolNumPages{Distill}{}{}{}.
\newblock
\APACrefnote{https://distill.pub/selforg/2021/textures}
\newblock
\begin{APACrefDOI} \doi{10.23915/distill.00027.003} \end{APACrefDOI}
\PrintBackRefs{\CurrentBib}

\bibitem [\protect \citeauthoryear {%
Pearson%
}{%
Pearson%
}{%
{\protect \APACyear {1993}}%
}]{%
pearson1993}
\APACinsertmetastar {%
pearson1993}%
\begin{APACrefauthors}%
Pearson, J\BPBI E.%
\end{APACrefauthors}%
\unskip\
\newblock
\APACrefYearMonthDay{1993}{}{}.
\newblock
{\BBOQ}\APACrefatitle {Complex patterns in a simple system} {Complex patterns in a simple system}.{\BBCQ}
\newblock
\APACjournalVolNumPages{Science}{261}{5118}{189--192}.
\PrintBackRefs{\CurrentBib}

\bibitem [\protect \citeauthoryear {%
Purves%
, Paydarfar%
\BCBL {}\ \BBA {} Andrews%
}{%
Purves%
\ \protect \BOthers {.}}{%
{\protect \APACyear {1996}}%
}]{%
purves1996}
\APACinsertmetastar {%
purves1996}%
\begin{APACrefauthors}%
Purves, D.%
, Paydarfar, J\BPBI A.%
\BCBL {}\ \BBA {} Andrews, T\BPBI J.%
\end{APACrefauthors}%
\unskip\
\newblock
\APACrefYearMonthDay{1996}{}{}.
\newblock
{\BBOQ}\APACrefatitle {The wagon wheel illusion in movies and reality.} {The wagon wheel illusion in movies and reality.}{\BBCQ}
\newblock
\APACjournalVolNumPages{Proceedings of the National Academy of Sciences of the United States of America}{93 8}{}{3693-7}.
\PrintBackRefs{\CurrentBib}

\bibitem [\protect \citeauthoryear {%
Rafler%
}{%
Rafler%
}{%
{\protect \APACyear {2011}}%
}]{%
rafler2011}
\APACinsertmetastar {%
rafler2011}%
\begin{APACrefauthors}%
Rafler, S.%
\end{APACrefauthors}%
\unskip\
\newblock
\APACrefYearMonthDay{2011}{}{}.
\newblock
{\BBOQ}\APACrefatitle {Generalization of Conway's "Game of Life" to a continuous domain - SmoothLife} {Generalization of conway's "game of life" to a continuous domain - smoothlife}.{\BBCQ}
\newblock
\APACjournalVolNumPages{arXiv: Cellular Automata and Lattice Gases}{}{}{}.
\PrintBackRefs{\CurrentBib}

\bibitem [\protect \citeauthoryear {%
Randazzo%
, Mordvintsev%
, Niklasson%
, Levin%
\BCBL {}\ \BBA {} Greydanus%
}{%
Randazzo%
\ \protect \BOthers {.}}{%
{\protect \APACyear {2020}}%
}]{%
randazzo2020}
\APACinsertmetastar {%
randazzo2020}%
\begin{APACrefauthors}%
Randazzo, E.%
, Mordvintsev, A.%
, Niklasson, E.%
, Levin, M.%
\BCBL {}\ \BBA {} Greydanus, S.%
\end{APACrefauthors}%
\unskip\
\newblock
\APACrefYearMonthDay{2020}{}{}.
\newblock
{\BBOQ}\APACrefatitle {Self-classifying MNIST Digits} {Self-classifying mnist digits}.{\BBCQ}
\newblock
\APACjournalVolNumPages{Distill}{}{}{}.
\newblock
\APACrefnote{https://distill.pub/2020/selforg/mnist}
\newblock
\begin{APACrefDOI} \doi{10.23915/distill.00027.002} \end{APACrefDOI}
\PrintBackRefs{\CurrentBib}

\bibitem [\protect \citeauthoryear {%
Rucker%
}{%
Rucker%
}{%
{\protect \APACyear {2003}}%
}]{%
rucker2003}
\APACinsertmetastar {%
rucker2003}%
\begin{APACrefauthors}%
Rucker, R.%
\end{APACrefauthors}%
\unskip\
\newblock
\APACrefYearMonthDay{2003}{}{}.
\newblock
{\BBOQ}\APACrefatitle {Continuous-Valued Cellular Automata in Two} {Continuous-valued cellular automata in two}.{\BBCQ}
\newblock
\APACjournalVolNumPages{New Constructions in Cellular Automata}{}{}{295}.
\PrintBackRefs{\CurrentBib}

\bibitem [\protect \citeauthoryear {%
Ryle%
}{%
Ryle%
}{%
{\protect \APACyear {1949}}%
}]{%
ryle1949}
\APACinsertmetastar {%
ryle1949}%
\begin{APACrefauthors}%
Ryle, G.%
\end{APACrefauthors}%
\unskip\
\newblock
\APACrefYear{1949}.
\newblock
\APACrefbtitle {The Concept of Mind} {The concept of mind}.
\newblock
\APACaddressPublisher{}{University of Chicago Press}.
\newblock
\begin{APACrefURL} \url{https://archive.org/details/conceptofmind032022mbp} \end{APACrefURL}
\PrintBackRefs{\CurrentBib}

\bibitem [\protect \citeauthoryear {%
Variengien%
, Nichele%
, Glover%
\BCBL {}\ \BBA {} Pontes-Filho%
}{%
Variengien%
\ \protect \BOthers {.}}{%
{\protect \APACyear {2021}}%
}]{%
variengien2021}
\APACinsertmetastar {%
variengien2021}%
\begin{APACrefauthors}%
Variengien, A.%
, Nichele, S.%
, Glover, T\BPBI E.%
\BCBL {}\ \BBA {} Pontes-Filho, S.%
\end{APACrefauthors}%
\unskip\
\newblock
\APACrefYearMonthDay{2021}{}{}.
\newblock
{\BBOQ}\APACrefatitle {Towards self-organized control: Using neural cellular automata to robustly control a cart-pole agent} {Towards self-organized control: Using neural cellular automata to robustly control a cart-pole agent}.{\BBCQ}
\newblock
\APACjournalVolNumPages{Innovations in Machine Intelligence (IMI)}{1}{}{1-14}.
\newblock
\begin{APACrefDOI} \doi{10.54854/imi2021.01} \end{APACrefDOI}
\PrintBackRefs{\CurrentBib}

\bibitem [\protect \citeauthoryear {%
von Neumann%
}{%
von Neumann%
}{%
{\protect \APACyear {1951}}%
}]{%
vonneumann1951}
\APACinsertmetastar {%
vonneumann1951}%
\begin{APACrefauthors}%
von Neumann, J.%
\end{APACrefauthors}%
\unskip\
\newblock
\APACrefYearMonthDay{1951}{}{}.
\newblock
{\BBOQ}\APACrefatitle {The General and Logical Theory of Automata} {The general and logical theory of automata}.{\BBCQ}
\newblock
\BIn{} L.~Jeffress\ (\BED), \APACrefbtitle {Cerebral Mechanisms in Behavior – The Hixon Symposium} {Cerebral mechanisms in behavior – the hixon symposium}\ (\BPG~1-31).
\newblock
\APACaddressPublisher{}{John Wiley \& Sons, New York, New York}.
\newblock
\begin{APACrefURL} \url{https://www.cs.unm.edu/~eschulte/classes/cs591-rpc/data/vonneumann1951-glta.pdf} \end{APACrefURL}
\PrintBackRefs{\CurrentBib}

\bibitem [\protect \citeauthoryear {%
von Neumann%
\ \BBA {} Burks%
}{%
von Neumann%
\ \BBA {} Burks%
}{%
{\protect \APACyear {1966}}%
}]{%
vonneumann1966}
\APACinsertmetastar {%
vonneumann1966}%
\begin{APACrefauthors}%
von Neumann, J.%
\BCBT {}\ \BBA {} Burks, A\BPBI W.%
\end{APACrefauthors}%
\unskip\
\newblock
\APACrefYearMonthDay{1966}{}{}.
\newblock
{\BBOQ}\APACrefatitle {Theory Of Self Reproducing Automata} {Theory of self reproducing automata}.{\BBCQ}
\newblock
\APACaddressPublisher{}{University of Illinois Press, Urbana and London}.
\newblock
\begin{APACrefURL} \url{https://web.archive.org/web/20220810124416/https://www.krusch.com/books/evolution/Theory_of_Self_Reproducing_Automata_Neumann1966.pdf} \end{APACrefURL}
\PrintBackRefs{\CurrentBib}

\bibitem [\protect \citeauthoryear {%
Wolfram%
}{%
Wolfram%
}{%
{\protect \APACyear {1984}}%
}]{%
wolfram1984}
\APACinsertmetastar {%
wolfram1984}%
\begin{APACrefauthors}%
Wolfram, S.%
\end{APACrefauthors}%
\unskip\
\newblock
\APACrefYearMonthDay{1984}{}{}.
\newblock
{\BBOQ}\APACrefatitle {Universality and complexity in cellular automata} {Universality and complexity in cellular automata}.{\BBCQ}
\newblock
\APACjournalVolNumPages{Physica D: Nonlinear Phenomena}{10}{1-2}{1--35}.
\PrintBackRefs{\CurrentBib}

\bibitem [\protect \citeauthoryear {%
Wulff%
\ \BBA {} Hertz%
}{%
Wulff%
\ \BBA {} Hertz%
}{%
{\protect \APACyear {1992}}%
}]{%
wulff1992}
\APACinsertmetastar {%
wulff1992}%
\begin{APACrefauthors}%
Wulff, N.%
\BCBT {}\ \BBA {} Hertz, J\BPBI A.%
\end{APACrefauthors}%
\unskip\
\newblock
\APACrefYearMonthDay{1992}{}{}.
\newblock
{\BBOQ}\APACrefatitle {Learning cellular automaton dynamics with neural networks} {Learning cellular automaton dynamics with neural networks}.{\BBCQ}
\newblock
\APACjournalVolNumPages{Advances in Neural Information Processing Systems}{5}{}{}.
\PrintBackRefs{\CurrentBib}

\end{thebibliography}

\end{document}